\documentclass[acmsmall]{acmart}
\usepackage{multirow}
\usepackage{afterpage}
\usepackage{lscape}
\usepackage{algorithm}
\usepackage{algpseudocode}

\AtBeginDocument{%
  }

\setcopyright{acmlicensed}
\copyrightyear{2018}
\acmYear{2018}
\acmDOI{XXXXXXX.XXXXXXX}

\acmJournal{TOSEM}
\acmVolume{37}
\acmNumber{4}
\acmArticle{111}
\acmMonth{8}

\begin{document}

\title{RepoRepair: Leveraging Code Documentation for Repository-Level Automated Program Repair}


\author{Zhongqiang Pan}
\email{602023720002@smail.nju.edu.cn}
\author{Chuanyi Li}
\authornote{Corresponding Author.}
\email{lcy@nju.edu.cn}
\author{Wenkang Zhong}
\email{zhongwenkang97@foxmail.com}
\author{Yi Feng}
\email{fy@nju.edu.cn}
\author{Bin Luo}
\email{luobin@nju.edu.cn}
\affiliation{%
  \institution{State Key Laboratory for Novel Software Technology, Nanjing University}
  \city{Nanjing}
  \country{China}
}
\author{Vincent Ng}
\email{vince@hlt.utdallas.edu}
\affiliation{
  \institution{Human Language Technology Research Institute, University of Texas at Dallas}
  \city{Richardson}
  \state{Texas}
  \country{USA}
}

\begin{abstract}
Automated program repair (APR) struggles to scale from isolated functions to full repositories, as it demands a global, task-aware understanding to locate necessary changes. Current methods, limited by context and reliant on shallow retrieval or costly agent iterations, falter on complex cross-file issues. To this end, we propose RepoRepair, a novel documentation-enhanced approach for repository-level fault localization and program repair. Our core insight is to leverage LLMs to generate hierarchical code documentation (from functions to files) for code repositories, creating structured semantic abstractions that enable LLMs to comprehend repository-level context and dependencies. Specifically, RepoRepair first employs a text-based LLM (e.g., DeepSeek-V3) to generate file/function-level code documentation for repositories, which serves as auxiliary knowledge to guide fault localization. Subsequently, based on the fault localization results and the issue description, a powerful LLM (e.g., Claude-4) attempts to repair the identified suspicious code snippets. Evaluated on SWE-bench Lite, RepoRepair achieves a 45.7\% repair rate at a low cost of \$0.44 per fix. On SWE-bench Multimodal, it delivers state-of-the-art performance with a 37.1\% repair rate despite a higher cost of \$0.56 per fix, demonstrating robust and cost-effective performance across diverse problem domains.
\end{abstract}




\received{20 February 2007}
\received[revised]{12 March 2009}
\received[accepted]{5 June 2009}

\maketitle

\section{Introduction}
The rapid evolution of large language models (LLMs) such as GPT-4\cite{gpt4} and DeepSeek-V3\cite{deepseek} has profoundly impacted software engineering, particularly in automating core development activities including code synthesis\cite{code_generation_1,code_generation_2,code_generation_3}, translation\cite{code_translation_1,code_translation_2,code_translation_3}, and repair\cite{repair-1, repair-2, repair-3}. While these advancements demonstrate proficiency in constrained settings, a critical gap persists when scaling to repository-level software maintenance. Here, the objective shifts from generating syntactically correct code snippets to executing semantically correct changes within a sprawling, interdependent codebase. This entails not merely parsing thousands of lines but discerning architectural intent, implicit contracts, and often non-textual cues from issue reports. Consequently, models must overcome two intertwined barriers: the prohibitive context window consumption of raw repository data and the semantic dispersion of fault-relevant information across multiple files and modalities, challenges that remain inadequately addressed by current paradigms.

To systematically evaluate the capability of LLMs in addressing real-world software engineering tasks, numerous benchmarks have been proposed. In program repair, Jimenez and Yang established two benchmark datasets in 2023 and 2024: \textbf{SWE-bench}\cite{swe-bench} and a multi-modal version \textbf{SWE-bench Multimodal}\cite{swe-bench-m}. Both datasets comprise real-world GitHub issue descriptions paired with the corresponding code repositories. Since the introduction of SWE-bench, the highest resolution rate in its most widely used subset, SWE-bench Lite, has increased from the initial 3\% to 60\% (excluding proprietary tools)\footnote{https://www.swebench.com/index.html (accessed September 2025)}. However, these tools demonstrate significantly inferior performance on SWE-bench Multimodal. For instance, the current state-of-the-art repair tool Agentless Lite\footnote{https://github.com/sorendunn/Agentless-Lite}, which achieves a 32.33\% resolution rate on SWE-bench Lite, experiences a notable performance drop to only 25.34\% on SWE-bench Multimodal. This performance gap arises primarily from two factors. First, many tools are designed without sufficient consideration for programming language generalization, resulting in core modules tightly coupled with Python (the back-end language in SWE-bench). For example, most of the tools provided by AutoCodeRover\cite{autocoderover} are highly based on Python-specific program analysis features and even require prior knowledge of the specific repositories in SWE-bench. Second, while some tools (e.g., SWE-Agent\cite{swe-agent} and Agentless\cite{agentless}) have decoupled from programming languages and can operate across languages, their utilization of repository information remains superficial, limited to directory structure trees and fragmented code snippets retrieved through search. This limitation is particularly problematic in SWE-bench Multimodal, where issues are often more complex and may span multiple files, making it difficult for these tools to accurately localize all suspicious code segments through superficial repository information alone. These limitations collectively point to a fundamental challenge: the lack of a structured, semantic representation of the repository that can effectively bridge the gap between high-level issue descriptions and low-level code changes.

To address the fundamental challenge of achieving deep, semantic understanding of repositories for automated program repair, we propose \textbf{RepoRepair}, a novel documentation-enhanced approach. Our work is motivated by a key insight: in software development, code documentation systematically abstracts implementation details, thereby serving as a crucial cognitive aid for program comprehension and maintenance\cite{code_doc_1,code_doc_2,code_doc_3}. We posit that leveraging LLMs to generate such structured documentation can create an effective intermediary representation, bridging the gap between high-level issue descriptions and the low-level code changes required for repair. While Luo and Ye demonstrated the utility of LLM-generated documentation for Python repositories\cite{repoagent}, we significantly extend and innovate upon this concept. Specifically, we introduce a novel hierarchical fault localization mechanism driven by two-tiered documentation: file-level documentation filters suspicious files, while subsequent function-level documentation pinpoints the exact suspicious classes or functions. This structured approach allows our method to overcome the superficial repository utilization of prior tools. \textbf{RepoRepair} operates in two stages: (1) In the localization stage, the hierarchical documentation is exclusively utilized to achieve precise fault localization; (2) In the repair stage, we introduce a novel, context-aware pipeline. This pipeline first prunes the localized code to construct a dependency-preserving repair context, then generates minimal diff-formatted patches, and finally validates them through an iterative, combinatorial testing strategy. This cohesive two-stage design enables that the deep understanding gained during localization is effectively translated into correct and reliable code changes, maximizing repair accuracy while maintaining strict cost control.

To evaluate the efficacy of \textbf{RepoRepair}, we conduct comprehensive experiments on two benchmark suites: the widely-adopted SWE-bench Lite and the more challenging SWE-bench Multimodal. Our experimental pipeline involves: (i) parsing source files using the open-source tree-sitter library\footnote{https://github.com/tree-sitter/tree-sitter}; (ii) generating hierarchical documentation via LLM; and (iii) processing issue descriptions (including multimodal content) during localization and repair. On SWE-bench Lite, \textbf{RepoRepair} achieves a competitive repair rate of 45.7\% at an average cost of just \$0.44 per fix, accompanied by a high file-level fault localization accuracy of 77.6\%. The superiority of our \textbf{RepoRepair} becomes even more pronounced on the complex SWE-bench Multimodal benchmark. \textbf{RepoRepair} attains a state-of-the-art repair rate of 37.1\% with excellent cost efficiency (\$0.56 per fix). Crucially, it achieves 59.8\% accuracy in file-level fault localization—representing a nearly 30\% absolute improvement over the prior best tool, Agentless Lite. This dramatic improvement in localization directly underpins the superior repair performance, demonstrating that our method excels precisely where prior approaches struggle: in comprehending and resolving complex, cross-file issues. These results collectively validate the generalizability, effectiveness, and practical efficiency of our documentation-enhanced framework across diverse programming environments and problem complexities.

The main contributions of this work are threefold:

\begin{itemize}
  \item \textbf{A Documentation-Enhanced Repair Framework}: We propose RepoRepair, a novel agent-free framework that (1) leverages hierarchical code documentation to significantly improve LLM-based fault localization from issue descriptions, and (2) implements an effective repair pipeline featuring combinatorial patch validation and temperature-based iterative generation to ensure robust patch generation.
  
  \item \textbf{Exhaustive Performance Benchmarking}: The systematic comparison with established baselines on both the SWE-bench and SWE-bench Multimodal benchmarks, evaluating not only final resolution rates but also quantifying differences in accuracy of file retrieval and file localization in the stage of localization.
  
  \item \textbf{Available Artifacts}: To facilitate reproducibility and future work, we publicly release all experimental artifacts (generated code documentation, source code, and results) in the Github repository: https://github.com/ZhongQiangDev/RepoRepair.
\end{itemize}

The rest of the paper is organized as follows. Section~\ref{sec:related} reviews related work. Section~\ref{sec:reporepair} details the design of RepoRepair. Section~\ref{sec:setup} describes the experimental setup, followed by Section~\ref{sec:results} which presents the experimental results. Section~\ref{sec:threats} discusses threats to validity, and Section~\ref{sec:limitations} outlines limitations. Finally, Section~\ref{sec:conclusion} concludes the paper and suggests directions for future work.

\section{Related Work}
\label{sec:related}
\subsection{Repository-level Automated Program Repair}

In the field of automatic program repair (APR), Defects4J\cite{defects4j} has long been considered the standard benchmark. However, with the introduction of SWE-bench benchmark by Jimenez and Yang\cite{swe-bench} in 2024, the focus of research has progressively shifted from traditional bug-fixing tasks at the function level to more challenging repository-level program repair. Crucially, many high-performing APR techniques in Defects4J, such as AlphaRepair\cite{alpharepair} and RepairLLaMA\cite{repairllama}, are exclusively designed to repair identified functions within known buggy files. These architectures are inadequate for repository-level program repair, which requires not only patch generation but also accurate file localization across the entire repository, substantially increasing task complexity.

Early explorations in repository-level APR often decoupled the localization and repair challenges. On the localization front, works like FuseFL\cite{fusefl} integrated traditional fault localization (e.g., SBFL) with LLMs to generate reasoning, while LLMAO\cite{llmao} fine-tuned code models for direct line-level localization. Conversely, on the repair front, methods like RLCE (Repository-Level Context Extraction)\cite{rlce} focused on retrieving relevant code snippets to enhance LLM-based patch generation, demonstrating improved performance on their self-constructed RepoBugs benchmark. These approaches, however, typically assumed that relevant files were pre-identified or operated under less stringent end-to-end constraints.
The advent of the SWE-bench benchmark, which pairs real-world GitHub issues with full repositories, formalized the need for truly integrated systems that perform both localization and repair from scratch. The seminal work by Jimenez and Yang\cite{swe-bench} itself proposed a retrieval-and-repair baseline, though initial results were low (3\% on SWE-bench Lite), highlighting the difficulty. This challenge spurred the development of more sophisticated methods, which can be broadly categorized into agent-based and agent-free paradigms.

With the rise of agent frameworks, Jimenez and Yang introduced SWE-Agent\cite{swe-agent}, the first agent-based repository-level APR approach. SWE-Agent employs an Agent-Computer Interface (ACI) to navigate repositories, edit files, and run tests. Concurrently, Zhang et al. proposed AutoCodeRover\cite{autocoderover}, which extracts structural specifications to guide the agent. Following this, multi-agent frameworks emerged: CodeR\cite{coder} employs specialized agents collaborating via task graphs, while SWE-Search\cite{swe-search} uses Monte Carlo Tree Search (MCTS)\cite{mcts} for dynamic planning. Recently, more performant agent-based approaches such as OpenHands\cite{openhands}, OpenHands-Versa\cite{openhands-2}, and ExpeRepair\cite{experepair} have demonstrated superior capabilities. However, agent-based approaches suffer from inherent limitations: they may uncritically accept misleading feedback and incur high computational costs due to excessive interaction rounds.

To mitigate these issues, Xia and Deng proposed Agentless\cite{agentless}, an agent-free APR method that adopts a streamlined two-stage localization-repair pipeline without requiring LLMs to decide actions or use complex tools. Surprisingly, Agentless matches or surpasses many agent-based tools in performance while drastically reducing costs, demonstrating the viability of agentless methods. More recently, GUIRepair\cite{guirepair} was specifically proposed for the SWE-bench Multimodal benchmark. It introduces dedicated Image2Code and Code2Image modules to process visual information (e.g., screenshots) and enhance repair accuracy for front-end issues, further advancing the agent-free paradigm.
While methods like Agentless and GUIRepair rely on raw code retrieval or visual processing, our proposed \textbf{RepoRepair} leverages hierarchically generated code documentation to capture semantic relationships across files, enabling more accurate localization and repair—particularly for issues requiring global repository understanding. It strengthens semantic representation through structured documentation, offering a distinct advancement in the agent-free paradigm.

\subsection{Automated Code Documentation Generation}

In software engineering, early automated documentation generation approaches\cite{doc_generation_1, doc_generation_2, doc_generation_3, doc_generation_4, doc_generation_5, doc_generation_6} are designed to alleviate developers' maintenance burdens by producing descriptive summaries for source code. However, these approaches suffered from a critical limitation: they focused solely on summarizing isolated code snippets while neglecting dependency relationships within broader repository-level contexts, often resulting in poor summarization.

To address this limitation, Luo and Ye proposed RepoAgent\cite{repoagent}, an agent-based automated code documentation framework designed for proactive generation, maintenance, and updating of code documentation in Python repositories. RepoAgent integrates three core modules: Global Structure Analysis (to capture repository-wide dependencies), Documentation Generation, and Documentation Update. Through blind preference testing, RepoAgent’s generated documentation is shown to be favored over human-written documentation by evaluators. Similarly, Yang et al. introduced DocAgent\cite{docagent}, a multi-agent system that tackles the same limitation by processing code in topological order and emulating human workflows. DocAgent employs five specialized agents, Reader, Searcher, Writer, Verifier, and Orchestrator, to collaboratively produce high-quality code documentation for Python Repository. 

While these works demonstrate the viability of automated, repository-aware documentation, they are primarily designed as general-purpose documentation tools. In contrast, our work repurposes and significantly adapts this paradigm for a specific downstream task: automated program repair. Building upon the foundational idea of LLM-generated documentation, we design a prompt-based generation approach that produces hierarchical documentation (function-level and file-level) explicitly optimized for fault localization. This task-driven design ensures the documentation captures the precise semantic and dependency information needed to guide repair, directly enhancing the effectiveness of our pipeline while improving scalability and cross-lingual adaptability.

\section{RepoRepair}
\label{sec:reporepair}

\begin{figure*}[htb]
\centerline{\includegraphics[width=\textwidth]{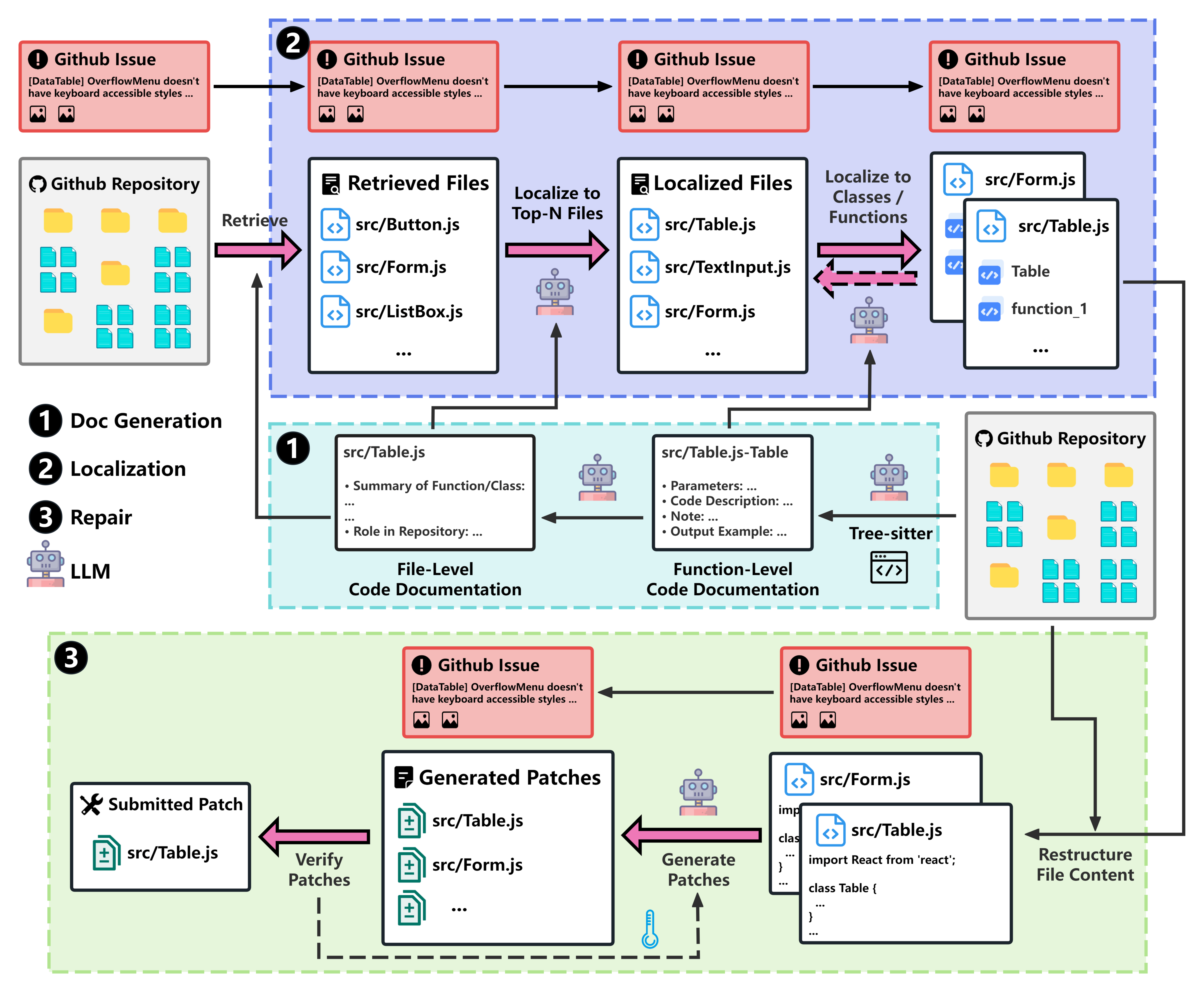}}
\caption{Overview of RepoRepair.}
\label{figure1}
\end{figure*}

Figure \ref{figure1} illustrates the overview of our proposed approach, RepoRepair, which consists of three key stages: \textbf{Documentation Generation}, \textbf{Localization}, and \textbf{Repair}. In the stages of \textbf{Documentation Generation}, we employ LLM to generate function-level and file-level code documentation for all code files in the target repository. In the stage of \textbf{Localization}, we first perform text-based retrieval to identify files whose file-level code documentation exhibits high semantic similarity to the issue description. Next, based on the issue description and the file-level code documentation we generated, we let LLM localize and rank the suspicious files that are most likely to need modification to solve the issue. Since not all content in these files may need editing, we further let LLM localize the functions/classes that need to be modified in each file based on the issue description and the function-level code documentation. In the stage of \textbf{Repair}, based on the functions/classes localized in each file, we first delete the contents (code and comments) of the remaining functions/classes that do not need to be modified and prune the content of the file for a minimal repair context. Then, we let LLM modify the content of each file and generate a diff format patch to attempt to solve the issue existing in the issue description. Finally, we verify these generated patches and select the most plausible patch of single or multiple files as the final submission. The complete workflow of RepoRepair is formalized in Algorithm~\ref{alg:reporepair}, and each stage will be described in detail in the following subsections.

\begin{algorithm}[htb]
\caption{RepoRepair}
\label{alg:reporepair}
\begin{algorithmic}[1]
\Require Code repository $\mathcal{R}$, issue description $\mathcal{I}$
\Ensure Repair patch $\mathcal{P}^*$ or $\emptyset$

\State \textbf{Stage 1: Documentation Generation}
\State $\mathcal{D} \gets \text{GenerateDocumentation}(\mathcal{R})$

\State \textbf{Stage 2: Localization}
\State $\mathcal{Q} \gets \text{Textualize}(\mathcal{I})$ \Comment{Convert multimodal issue to text}
\State $\mathcal{F}_s \gets \text{LocalizeFiles}(\mathcal{Q}, \mathcal{D})$ 
\State $\mathcal{N}_s \gets \text{LocalizeCodeUnits}(\mathcal{F}_s, \mathcal{Q})$

\If{$\mathcal{F}_s = \emptyset$ \textbf{or} $\mathcal{N}_s = \emptyset$}
    \State \Return $\emptyset$ \Comment{Cannot localize the issue}
\EndIf

\State \textbf{Stage 3: Repair}
\For{$T \in [0.0, 0.9]$ \textbf{step} 0.1}
    \State $\mathcal{P}_{\text{all}} \gets \text{GeneratePatches}(\mathcal{F}_s, \mathcal{N}_s, \mathcal{Q}, T)$ \Comment{Generate $C(n,k)$ combinatorial patches}
    \State $\mathcal{P}_{\text{valid}} \gets \text{ValidatePatches}(\mathcal{P}_{\text{all}}, \mathcal{R})$
    
    \If{$\mathcal{P}_{\text{valid}} \neq \emptyset$}
        \State $\mathcal{P}^* \gets \arg\min_{P \in \mathcal{P}_{\text{valid}}} \text{NumModifiedFiles}(P)$ 
        \State \Return $\mathcal{P}^*$  \Comment{Return the patch for final submission}
    \EndIf
\EndFor

\State \Return $\emptyset$ \Comment{Repair failed}
\end{algorithmic}
\end{algorithm}

\subsection{Documentation Generation}

As previously discussed, code documentation systematically describes implementation logic, significantly aiding developers in comprehending functionality, a critical factor for the program repair task that requires deep code understanding. However, manually creating high-quality code documentation for the entire repository requires substantial time, labor, and financial resources\cite{code_doc_3}. More critically, as the repository evolves, code documentation requires synchronous updates, yet most projects lack sufficient resources or incentives to maintain real-time documentation freshness.

To overcome this barrier and enable scalable, automated code understanding, we adopt a strategy of generating comprehensive, hierarchical code documentation for the entire repository. This foundational step, while computationally intensive, is critical for two reasons. First, it provides a complete and consistent knowledge base for the subsequent localization stage, ensuring that no potentially relevant code is overlooked during cross-file analysis—a common challenge in large-scale repositories. Second, by generating documentation incrementally as the repository evolves (detailed later in this section), we amortize the initial generation cost over time, aligning with real-world development workflows where issues are addressed sequentially. This approach is inspired by RepoAgent's methodology\cite{repoagent}, shifting from general repository comprehension to a targeted, hierarchical documentation system (function-level and file-level) explicitly designed to guide precise fault localization and repair.

\textbf{Generate function-level documentation}. Before generating documentation, we first convert all source files into a uniform intermediate representation. Using Tree-sitter and its multi-language grammars, we parse code into ASTs to systematically extract key program constructs—functions, classes, imports, and calls—while abstracting away syntactic particulars. This normalized representation decouples documentation generation from language-specific syntax, allowing consistent semantic extraction across Python, JavaScript, TypeScript, and other supported languages. From the parsed ASTs, we then extract critical information for all functions/classes, including name, signature, context, dependencies, etc. For anonymous functions in JavaScript/TypeScript, such as arrow functions or function expressions, we assign standardized names in the format function\_X, where X is a unique number. This process guarantees that the generated documentation contains complete structural metadata, providing a reliable foundation for subsequent fault localization.

\begin{figure*}[tbp]
\centerline{\includegraphics[width=\textwidth]{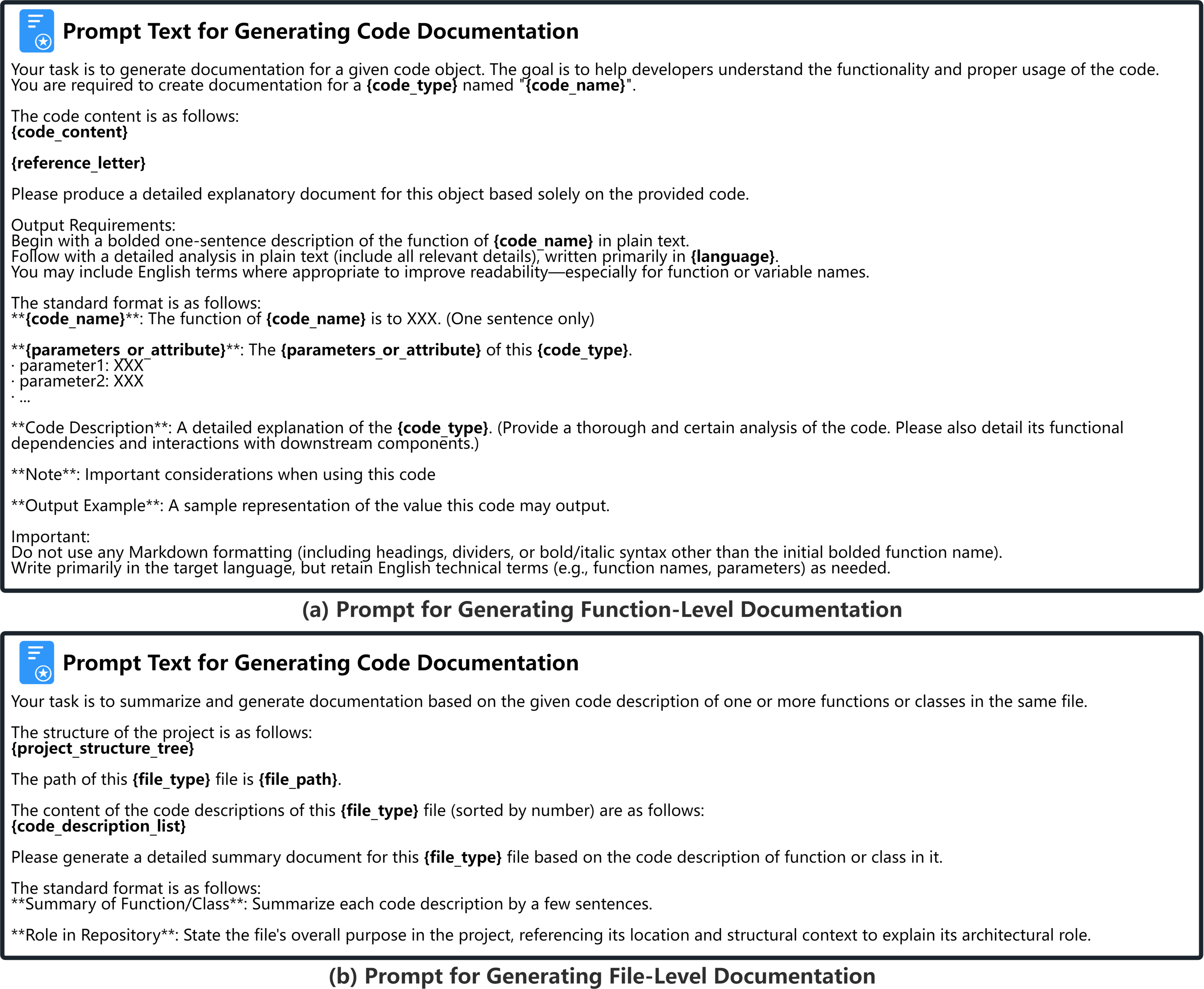}}
\caption{Prompt Texts for Generating Code Documentation.}
\label{figure2}
\end{figure*}

Subsequently, to generate high-quality documentation, we construct a prompt (Figure \ref{figure2}(a)) that includes the complete source code of the target function or class alongside its extracted metadata. We provide the full implementation rather than a summary or snippets to ensure the LLM has access to all necessary contextual details (e.g., internal logic, variable usage, and control flow) for producing accurate and informative descriptions. This input is then processed by the LLM to generate standardized code documentation containing: parameter introductions, code descriptions, usage notes, and output examples, formatted as shown in Figure \ref{figure1}.

\textbf{Generate file-level documentation}. After generating function-level code documentation for all files, we proceed to generate file-level code documentation. This higher-level abstraction shifts focus from implementation details to summarizing each file's overall purpose and architectural role within the repository—information crucial for the initial file retrieval and ranking steps in the localization stage.

This process takes two key inputs: (1) all function-level code documentation for the file, and (2) the project's structural tree, which provides crucial context about the file's position and relationships within the repository hierarchy. We input these elements to the LLM using a prompt (Figure \ref{figure2}(b)) that instructs it to generate comprehensive documentation containing: (1) a concise summary of each function/class's functionality and (2) a high-level description of the file's architectural role and dependencies within the project. The resulting documentation, formatted as shown in Figure \ref{figure1}, captures both the implementation specifics and the architectural significance of each file, enabling more effective repository-level understanding for subsequent localization and repair tasks.

\subsection{Localization}

To implement bug fixes or new features, accurate localization in the source code is fundamentally essential - without correct localization, correct modifications cannot be made. However, the challenge lies in the fact that the repository may contain hundreds or even thousands of code files, each potentially comprising thousands of lines of code, while the actual edit location might be just a few lines within a specific function/class. To address this challenge, we employ a hierarchical three-step localization pipeline: (i) retrieving files with high relevance to the issue, (ii) identifying suspicious files among them, and (iii) localizing the specific functions/classes related to the issue in each suspicious file.

\textbf{Retrieve most relevant files}. Given that the repository typically contains hundreds of code files while only a few may need modification, directly inputting all corresponding source code would result in prohibitively long contexts. To address this, we employ file-level code documentation as a proxy for retrieval instead of raw code. This design choice is motivated by two key considerations: (i) Efficiency: Documentation provides a condensed, semantically rich representation of each file's purpose and key functionalities, drastically reducing token count compared to full source code. (ii) Alignment with Natural Language Queries: Since issue descriptions are written in natural language, matching them against similarly structured documentation (rather than raw syntax) facilitates more accurate semantic retrieval. Specifically, we first vectorize all documentation using an embedding model and construct indices, then use the issue description as a query to retrieve the most relevant files.

\begin{figure}[tbp]
\centerline{\includegraphics[width=0.9\linewidth]{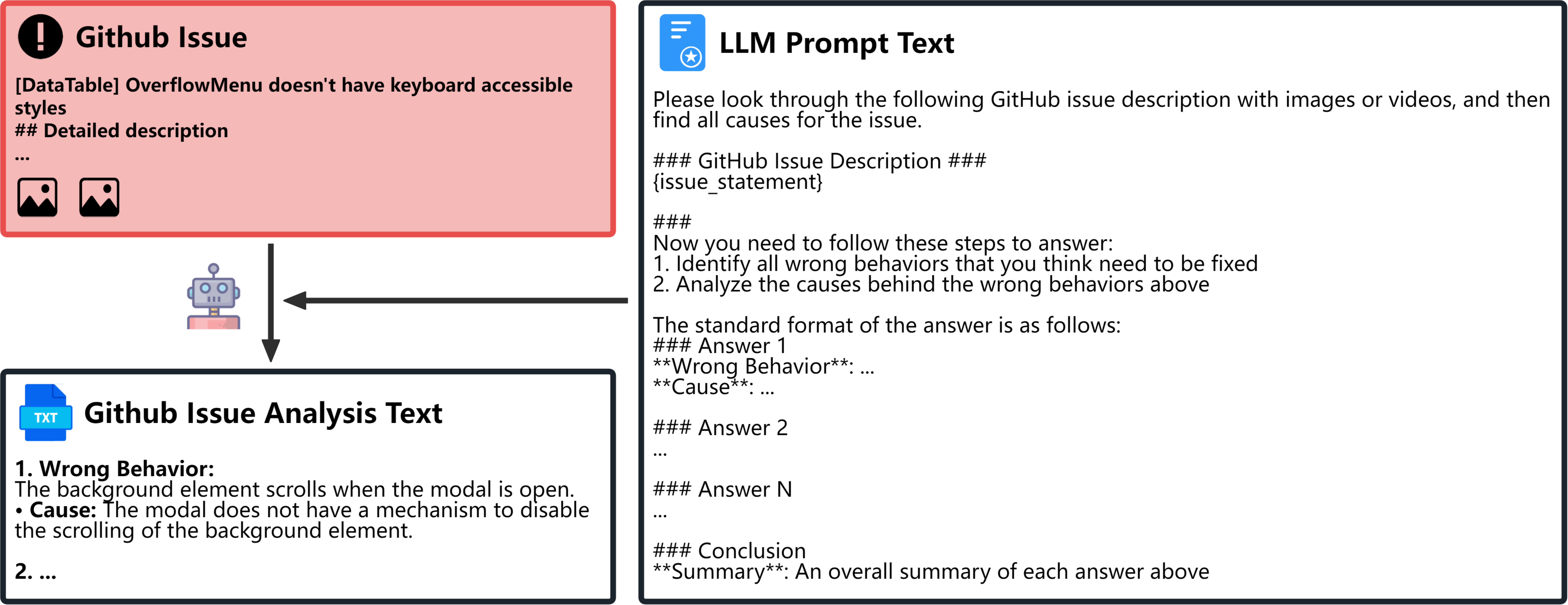}}
\caption{Multimodal Issue Description Processing for Retrieval.}
\label{figure3}
\end{figure}

Inspired by previous multimodal retrieval work\cite{retrival-2}, our approach strategically unifies all query-side information (including visual details such as screenshots) into textual representations before matching with the text-only code documentation database. This design offers two key advantages: (i) It avoids reliance on general-purpose multimodal embeddings (e.g., CLIP\cite{clip}) that may not be optimized for capturing front-end-specific code-visual relationships; (ii) It shifts the complex cross-modal alignment task to the powerful LLM during query processing, eliminating the need for real-time multimodal similarity computation during retrieval, thus improving efficiency. Specifically, for a multimodal issue description, we use a multimodal LLM to generate a comprehensive textual analysis of the issue, describing all possible error behaviors and root causes (shown in Figure \ref{figure3}). This generated text then serves as the retrieval query.

\textbf{Localize to suspicious files}. To identify suspicious files from the retrieved set, we continue to provide LLM with file-level code documentation rather than the complete source code. This is a deliberate continuation of our efficiency-focused design. While providing the full code might offer marginally more context, the accompanying surge in token volume would severely limit the number of files we can practically present to the LLM for analysis within budget constraints. The documentation, having been distilled to capture architectural roles and key functionalities, provides sufficient high-level context for the LLM to reason about which files are most likely to contain the fault, as demonstrated by its use of Chain-of-Thought (COT)\cite{cot} reasoning in our example. When analyzing the issue "carbon-design-system/carbon-3918" where "OverflowMenu keyboard navigation fails inside DataTable", the COT process first \textbf{\textit{identifies the core error behavior}} ("no visual indicators for keyboard navigation in DataTable-embedded OverflowMenu"), then \textbf{\textit{analyzes root causes}} such as missing keyboard event handlers in "DataTable.js" or conflicting focus-state styles in "OverflowMenu.scss", before finally \textbf{\textit{correlating these with relevant files}}. This approach ensures that localization decisions are based on a deep understanding of component interaction rather than superficial keyword matching.

\textbf{Localize to related functions/classes}. After identifying suspicious files, we further localize the specific functions/classes requiring modification. Here, we again use function-level code documentation as the input context. We deliberately avoid using more concise "file skeletons" (e.g., just function signatures) as done in some prior work\cite{agentless}, because such skeletons fail to capture anonymous functions and lack implementation context, which are critical for accurate localization in JavaScript/TypeScript codebases. Conversely, providing the full source code of all functions would be computationally prohibitive. Function-level documentation strikes an optimal balance: it is far more compact than full code, yet rich enough in semantic detail (describing logic, parameters, and behavior) to enable precise pinpointing of the relevant code units.

Notably, the function-level localization phase also serves as an essential verification mechanism for the preceding file-level localization. If the process fails to identify any relevant functions/classes within the suspicious files, this indicates potential inaccuracies in the previous file localization. In such cases, our approach automatically discards the current set of suspicious files and re-executes the file localization process to obtain a more reliable result.

\subsection{Repair}\label{repair}

\begin{figure}[tbp]
\centerline{\includegraphics[width=\linewidth]{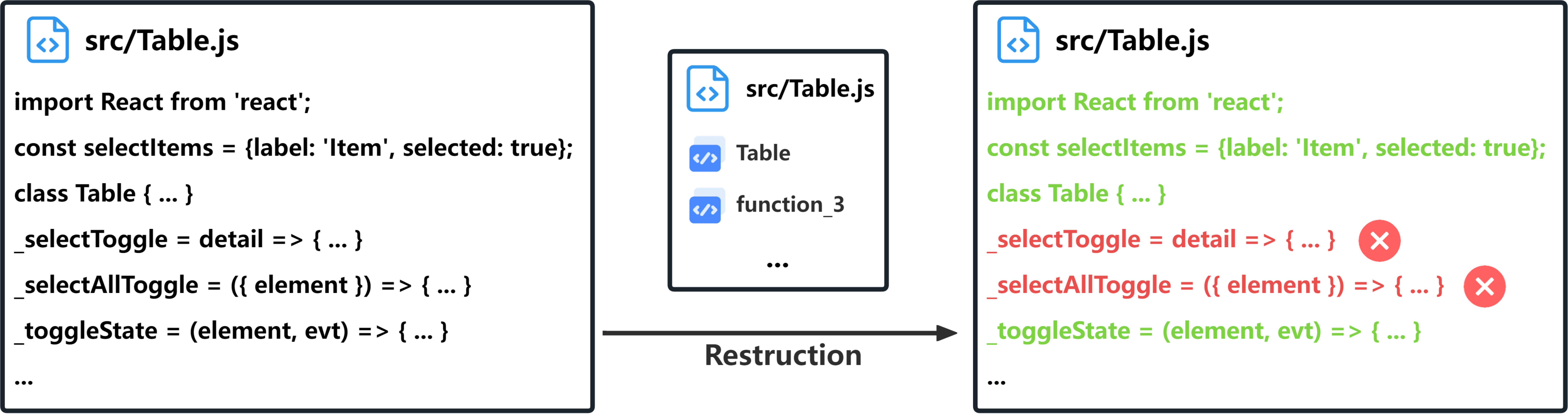}}
\caption{Pruning a Code File.}
\label{figure4}
\end{figure}

During the repair stage, our final goal is to generate correct patches that effectively resolve the issue. Constructing an effective repair context presents a dual challenge: it must contain enough semantic information to guide correct code modifications, yet remain concise to avoid overwhelming the LLM with irrelevant content. Traditional approaches often resort to arbitrary truncation or retain entire files, both of which can obscure critical dependencies or introduce noise.

To address this, we implement a dependency‑aware code pruning strategy (illustrated in Figure \ref{figure4}), which moves beyond simple context‑window truncation by selectively retaining code based on semantic relevance and static references. This design reflects a deliberate trade‑off that is particularly effective in front‑end and multi‑file repair scenarios. Specifically, we preserve three categories of content: (i) all localized functions or classes identified as primary edit targets; (ii) essential file‑level constructs such as import/require statements and global variable declarations that provide structural scaffolding; and (iii) any non‑localized functions or classes that are statically referenced by the preserved repair targets, ensuring intra‑file dependency integrity.

Crucially, our strategy incorporates a dependency-aware preservation rule to address data and control flow concerns: we retain not only localized code but also any unlocalized functions or classes that are statically referenced by it within the same file. This ensures that immediate, in‑file dependencies are maintained, directly mitigating the risk of overlooking relevant context. The counterpart is the aggressive removal of all other code—specifically, implementations of unlocalized and unreferenced units. We explicitly choose this trade‑off for efficiency and focus: while preserving full transitive dependencies could offer theoretical completeness, it would cause context explosion (especially in modular front‑end codebases), dilute the LLM’s attention with indirectly related code, and nullify the efficiency gains of our targeted localization. Our evaluation validates that for typical UI‑based issues, this triad—localized code, its direct dependencies, and file‑level scaffolding—provides a sufficiently rich yet focused context, maximizing repair accuracy while strictly controlling computational cost.

\textbf{Generate patch}. We sequentially input the pruned code files into LLM, which then generates diff-formatted patches based on the issue description for each file, rather than producing complete replacement code segments. Specifically, these diff patches consist of Search/Replace edit pairs\footnote{https://aider.chat/} containing two key components: (i) the search segment (original code to be modified) and (ii) the replacement segment (new code). To apply such patches to the original code files, one simply matches the search segment and substitutes it with the corresponding replacement segment. This kind of diff patch offers three significant advantages over complete code generation: first, it constrains LLM to focus on minimal, targeted modifications rather than complete rewrites; second, it dramatically improves cost-efficiency by reducing output size and complexity; and third, it enhances reliability and accuracy by significantly lowering the probability of hallucinations that often occur when generating large code blocks from scratch.


\textbf{Verify patch}. To ensure the correctness and reliability of the generated patches, we adopt a systematic, iterative validation pipeline that integrates traditional software engineering testing practices. The workflow, detailed in Algorithm~\ref{alg:reporepair}, proceeds as follows:

1. Patch Collection per Iteration: For a given issue, after the LLM generates candidate patches (one per localized file), we do not validate them individually. Instead, we aggregate patches into logically grouped sets based on the number of concurrently modified files. Specifically, within each validation iteration, we collect all possible patch combinations across the $n$ localized files, starting from single-file patches ($C(n,1)$ sets), followed by two-file combinations ($C(n,2)$ sets), up to the full set of all $n$ patches applied together ($C(n,n)$ set). This comprehensive combinatorial approach allows us to evaluate not only isolated changes but also potential synergistic or conflicting modifications across multiple files.

2. Batch Validation: All patch combinations collected in a single iteration are validated in parallel against the repository's existing test suite, performing regression testing \cite{test}. Among those combinations that pass all tests, we employ a minimal-change criterion: select the patch combination that modifies the fewest number of files. Formally, we choose $\mathcal{P}^* = \arg\min_{P \in \mathcal{P}{\text{pass}}} |\text{Files}(P)|$, where $\mathcal{P}{\text{pass}}$ denotes the set of all patch combinations that pass validation. This preference for smaller changes aligns with software maintenance best practices, favoring localized fixes that reduce the risk of unintended side effects. If no patch combination passes the tests in the current batch, the entire batch is discarded.

3. Temperature-based Iteration: If no patch combination from the current batch passes all tests, we initiate a new repair iteration. In each subsequent iteration, we increment the LLM's sampling temperature by 0.1 (starting from 0) to introduce controlled variation in the generated patches, encouraging diverse solutions. This loop continues until either a valid patch combination is found or the temperature reaches 0.9. The iterative, temperature-annealing strategy balances exploration of the repair space with the need for deterministic convergence.

\section{Experimental Setup}
\label{sec:setup}

\subsection{Dataset}

To rigorously assess RepoRepair’s efficacy and generalizability, we conduct experiments on two benchmarks that represent distinct facets of real-world repair: SWE-bench Lite and SWE-bench Multimodal. This dual evaluation tests whether our documentation-enhanced approach can handle both traditional code-centric bugs and emerging challenges in modern web development that require understanding visual UI behavior. SWE-bench Lite consists of 323 issues (23 in the 'dev' subset and 300 in the 'test' subset) drawn from Python repositories, featuring purely textual problem descriptions and predominantly single-file modifications. In contrast, SWE-bench Multimodal comprises 619 issues (102 in the 'dev' subset and 517 in the 'test' subset) from JavaScript/TypeScript repositories, incorporating multimodal information (both textual and visual elements) in issue descriptions and frequently involving more complex, cross-file changes.

The key distinctions between these two benchmarks lie in programming language diversity, issue description modalities, and repair complexity. While SWE-bench Lite focuses on Python-based issues with text-only descriptions, SWE-bench Multimodal extends the evaluation to JavaScript/TypeScript environments with rich multimodal context. Empirical observations confirm that approaches achieving high performance on SWE-bench Lite consistently demonstrate significantly lower results on SWE-bench Multimodal, indicating its inherently greater challenge level.

Furthermore, given that SWE-bench Lite was introduced over a year ago and the rapid iteration of large language models raises inevitable data leakage concerns, evaluation limited solely to SWE-bench Lite becomes increasingly insufficient. Therefore, conducting comprehensive evaluation on both benchmarks—particularly the more recent and challenging SWE-bench Multimodal—provides a more rigorous and practical assessment of program repair capabilities.


\subsection{Baselines}



To ensure a fair and comprehensive evaluation, we establish a systematic set of criteria for selecting baseline tools. Our selection aims to capture the state-of-the-art across three critical dimensions: generalizability, architectural relevance, and reproducibility.

\textbf{General Baselines}. To ensure our comparison reflects generalizability across benchmarks, we include all three tools that have been officially evaluated on both SWE-bench Lite and SWE-bench Multimodal: RAG\cite{swe-bench}, SWE-Agent\cite{swe-agent}, and Agentless\cite{agentless}. This ensures that our primary baselines are capable of handling both traditional code-only issues and emerging multimodal scenarios, providing a common ground for cross-dataset performance analysis.

\textbf{Architecturally Relevant Baselines}. For architectural relevance, we select Agentless Lite (the best-performing variant of Agentless) as our primary architectural baseline. This decision is based on its streamlined two-stage pipeline—first retrieving relevant files via embedding models, then iteratively generating repairs based on the top retrieved files—which shares key design principles with our approach (retrieval-augmented, context-aware repair) while maintaining computational efficiency. This focused comparison allows us to isolate and analyze the impact of our specific contributions, such as hierarchical documentation and code pruning, against a conceptually similar but implementationally distinct baseline.

\textbf{Benchmark-Specific Baselines}. To ensure comprehensiveness within each benchmark's ecosystem, we include all major, reproducible open-source tools specifically reported on each dataset. For SWE-bench Lite, this encompasses six additional tools: AutoCodeRover\cite{autocoderover}, OpenHands\cite{openhands}, PatchPilot\cite{patchpilot}, DARS\cite{dars}, ExpeRepair\cite{experepair}, SWE-Search\cite{swe-search}, and SpecRover\cite{specrover}. For SWE-bench Multimodal, we include two additional tools: OpenHands-Multimodal\cite{openhands-2} and GUIRepair\cite{guirepair}. This ensures our evaluation covers the full spectrum of published solutions in each domain, preventing selection bias and enabling a thorough assessment of capabilities across language-specific features, multimodal adaptation, and repair complexity.


\subsection{Detailed Settings of RepoRepair}

In the documentation generation stage of RepoRepair, we use DeepSeek-V3\cite{deepseek} (DeepSeek-V3-0324) to generate both function-level and file-level code documentation. To ensure deterministic output and factual accuracy, we set the sampling temperature to 0 while keeping all other parameters at their default values. For the localization and repair phases, we use a multimodal LLM Claude 4 (claude-4-sonnet-20250514), configuring the sampling temperature similarly to 0 and fixing the random seed to guarantee stable and reproducible output. For dynamic visual input (e.g., GIFs and videos) to the multimodal LLM, we extract key frames by computing the Structural Similarity Index (SSIM)\cite{ssim} between consecutive frames. This preprocessing step ensures efficient representation of temporal visual information while preserving critical content.

In the documentation generation stage, we treat all issues within a repository as representing its evolving state across different periods rather than independent projects. Our approach processes issues in strict ascending order of their GitHub-assigned IDs (which follow chronological sequence) and implements incremental code documentation generation. When code documentation of a previous issue exists, we only generate documentation for files that differ from the prior issue's state. This strategy offers two advantages: it optimizes token usage by avoiding redundant documentation of unchanged files and accurately mirrors real-world repository maintenance workflows. Although this mechanism may incur substantial token costs during initial stages, the average cost per issue gradually decreases as the number of resolved issues increases, resulting in significant long-term efficiency gains.

In the localization stage, our retrieval process uses sentence-transformers/all-mpnet-base-v2\footnote{https://huggingface.co/sentence-transformers/all-mpnet-base-v2} model as the embedding model to store vectorized representations in a FAISS\cite{faiss} database. For each issue description, we retrieve the top-50 most relevant files based on semantic similarity. Two practical nuances affect the actual number of files retrieved: (i) if duplicate files appear in the top-50 results, deduplication reduces the final count; (ii) when an issue description is processed through multiple queries, the union of results across queries may exceed 50 files. After retrieval, we proceed to localize the top-5 most suspicious files from the retrieved set and then localize all potentially relevant functions/classes within them, without quantity restrictions.

In the repair stage, we adopt an iterative validation approach as described in Section \ref{repair}. To balance repair exploration with computational efficiency, we set the LLM temperature to increase by 0.1 in each iteration until reaching 0.9. All patch combinations within a batch are validated in parallel against the repository's existing test suite, subject to a per-process timeout threshold of one minute. If the validation of any individual patch combination exceeds this limit, that specific combination is considered failed; the validation of other combinations in the batch proceeds independently.

\subsection{Evaluation Metrics}

For evaluation on SWE-bench Lite and SWE-bench Multimodal, we adopt the established metrics from prior work \cite{autocoderover, agentless, swe-bench-m}: (1) \textbf{\%Resolved} - the percentage of issues successfully fixed in the benchmark. Notably, for repository-level evaluation, all patches are validated via automated test suites; any patch that passes is considered correct, and the distinction of "plausible patch" is no longer used. (2) \textbf{Avg. \$Cost} - the average inference cost per tool run. Additionally, we introduce \textbf{\%Correct Localization} - the percentage of cases where the tool's localized files exactly match those modified in developer patches. We measure this metric at the stage of localization, considering localization to be correct if RepoRepair's localized file set fully contains all developer-patched files. It is a recall-oriented metric used in prior work\cite{agentless} to assess whether a tool can successfully identify all relevant fault locations, even at the potential cost of including some false positives. For baseline tools, we use officially reported test results from their repositories.

\section{Experimental Results}
\label{sec:results}


  
  


\subsection{Overall Performance}

Table \ref{table1} presents the overall performance metrics (\%Resolved and Avg. \$Cost) on the SWE-bench Lite benchmark across all methods, while Table \ref{table2} shows the corresponding results on the SWE-bench Multimodal benchmark. Tables \ref{table3} and \ref{table4} provide granular analyzes comparing RepoRepair with Agentless Lite and GUIRepair across both datasets, specifically examining repair success rates (\%Resolved) and localization accuracy (\%Correct Localization) across different repositories.

\begin{table}[tb]
\caption{Performance Comparison of Baseline Tools on SWE-bench Lite.}
\label{table1}
\centering
\begin{tabular}{llcc}
\toprule
Tool & LLM & \%Resolved & Avg. \$Cost \\
\midrule
RAG & Claude-3 Opus & 4.33\% & -\\
SWE-Agent & Claude-3.5 Sonnet & 23.00\% & 1.62 \\
SWE-Agent & Claude-4 Sonnet & 56.67\% & - \\
AutoCodeRover & GPT-4o & 30.67\% & 0.65 \\
Moatless Tools & Claude-3.5 Sonnet & 38.33\% & 0.17 \\
Agentless & GPT-4o & 27.33\% & 0.34 \\
Agentless Lite & O3 Mini & 32.33\% & 0.21 \\
OpenHands & CodeAct v2.1 & 41.67\% & 2.14 \\
PatchPilot & Claude-3.5 Sonnet & 45.33\% & 0.97 \\
DARS & Claude-3.5 Sonnet & 47.00\% & 12.24 \\
ExpeRepair & Claude-3.5 Sonnet & 48.33\% & 2.07 \\ 
ExpeRepair & Claude-4 Sonnet & \textbf{60.33\%} & - \\ \hline
RepoRepair & \begin{tabular}[c]{@{}l@{}}Deepseek-V3 \& \\ Claude-4 Sonnet\end{tabular} & 45.67\% & 0.44 \\
\bottomrule
\end{tabular}
\end{table}

\begin{table}[tb]
\caption{Performance Comparison of Baseline Tools on SWE-bench Multimodal.}
\label{table2}
\centering
\begin{tabular}{llcc}
\toprule
Tool & LLM & \%Resolved & Avg. \$Cost \\
\midrule
RAG & GPT-4o & 6.00\% & 0.17\\
SWE-Agent & Claude-3.5 Sonnet & 12.19\% & 1.52 \\
SWE-Agent JS & Claude-3.5 Sonnet & 11.99\% & 3.11 \\
SWE-agent M & GPT-4o & 12.19\% & 2.94 \\
Agentless & Claude-3.5 Sonnet & 6.19\% & 0.42 \\
Agentless Lite & Claude-3.5 Sonnet & 25.34\% & 0.38 \\ 
OpenHands-Versa & Claude-4 Sonnet & 34.43\% & - \\ 
GUIRepair & o4-mini & 33.85\% & 0.36 \\ 
GUIRepair & o3 & 35.98\% & - \\ \hline
RepoRepair & \begin{tabular}[c]{@{}l@{}}Deepseek-V3 \& \\ Claude-4 Sonnet\end{tabular} & \textbf{37.14\%} & 0.56 \\
\bottomrule
\end{tabular}
\end{table}

\textbf{Performance Comparison with Baseline Tools}. Tables~\ref{table1} and~\ref{table2} present comprehensive comparisons of repair performance (\%Resolved) on SWE-bench Lite and Multimodal benchmarks, respectively. On SWE-bench Lite, RepoRepair achieves a competitive 45.67\% resolution rate with favorable cost efficiency (\$0.44 per issue). While ExpeRepair with Claude-4 Sonnet attains the highest performance (60.33\%), our approach outperforms several prominent baselines such as PatchPilot (45.67\% vs. 45.33\%), OpenHands (45.67\% vs. 41.67\%), and DARS (45.67\% vs. 47.00\%) at significantly lower cost. More importantly, on the more challenging SWE-bench Multimodal benchmark, RepoRepair establishes state-of-the-art performance with 37.14\% resolution, surpassing all existing tools including GUIRepair (35.98\%), OpenHands-Versa (34.43\%), and Agentless Lite (25.34\%). These results demonstrate that our RepoRepair achieves competitive performance while maintaining cost efficiency across diverse benchmarks.

\begin{figure}[t]
\centering
\includegraphics[width=\linewidth]{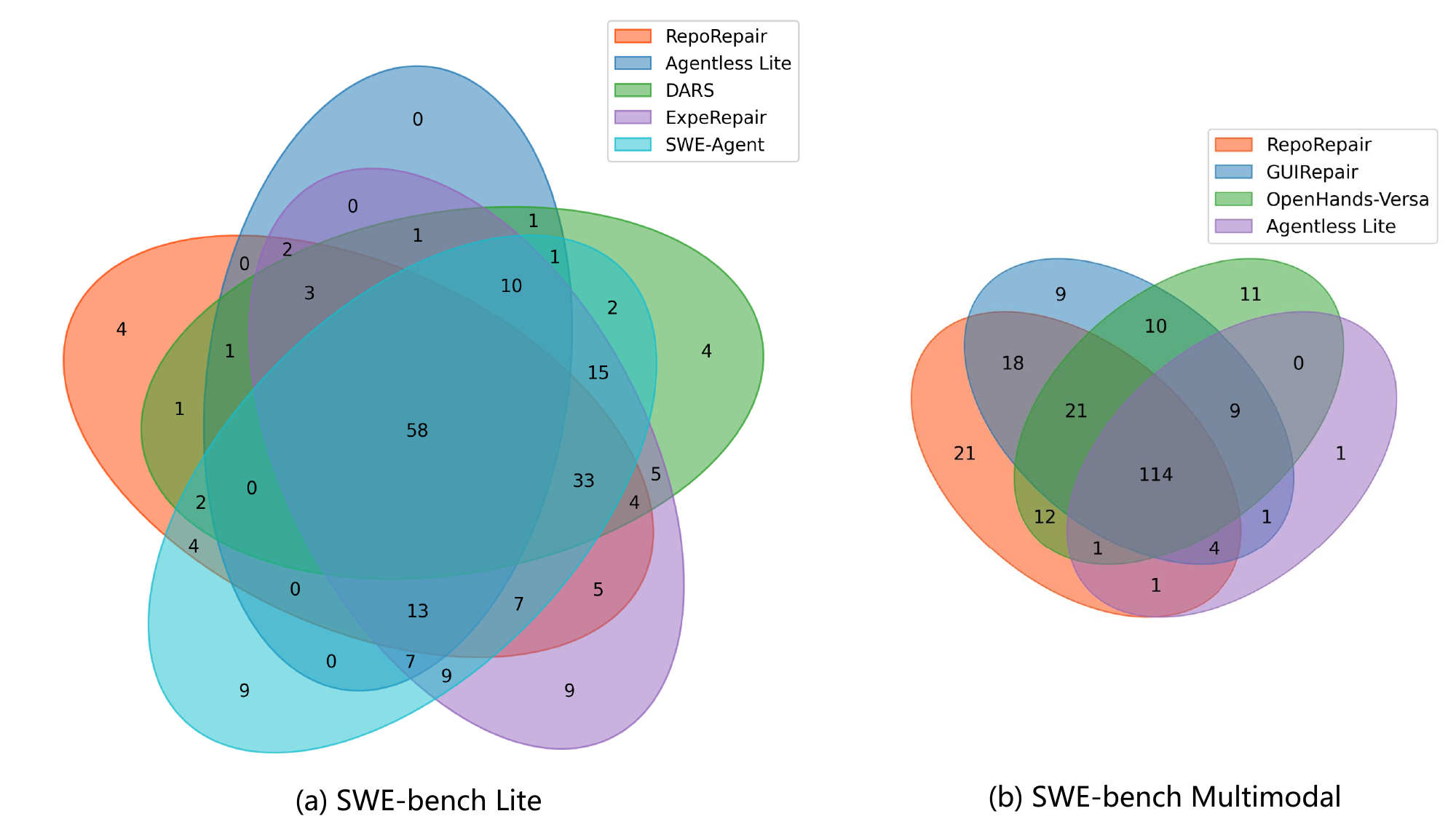}
\caption{Unique bug resolutions of top-performing tools on (left) SWE-bench Lite and (right) SWE-bench Multimodal.}
\label{figure5}
\end{figure}

\textbf{Unique Repair Capabilities}. Complementing the aggregate performance metrics, Figure \ref{figure5} provides a Venn diagram analysis that reveals RepoRepair's distinctive ability to resolve bugs that elude other top-performing tools. On SWE-bench Lite, RepoRepair uniquely fixes 4 bugs that all four comparison tools (Agentless Lite, DARS, ExpeRepair, SWE-Agent) fail to resolve. This unique capability is even more pronounced on SWE-bench Multimodal, where RepoRepair solves 21 bugs that GUIRepair, OpenHands-Versa, and Agentless Lite cannot fix. These uniquely resolved cases often involve issues with minimal file mention in descriptions, requiring deep semantic understanding of repository structure—a strength of RepoRepair. This analysis underscores that beyond competitive overall performance, RepoRepair addresses a valuable subset of challenging problems inaccessible to existing methods.

Furthermore, we conduct a comparative analysis of performance differences between RepoRepair, Agentless Lite, and GUIRepair on SWE-bench Multimodal. As shown in Table~\ref{table3}, RepoRepair achieves the highest overall performance (37.14\%) compared to GUIRepair (35.98\%) and Agentless Lite (25.34\%). Notably, the performance patterns vary significantly across different repository types. RepoRepair achieves particularly strong results in text/code-focused repositories: 'bpmn-io/bpmn-js' and 'eslint/eslint'. These repositories primarily involve code logic, API usage, and configuration issues—domains where RepoRepair effectively bridges issue descriptions to relevant code elements. In contrast, GUIRepair, which leverages visual information from screenshots for reproduction and validation, shows relative advantages in repositories with significant visual components. In 'GoogleChrome/lighthouse', its ability to process visual information provides tangible benefits for issues involving web rendering and visualization. Similarly, in syntax-focused repositories like 'PrismJS/prism' and 'prettier/prettier', GUIRepair's visual validation mechanism contributes to its performance. Despite GUIRepair's advantages in specific visual domains, RepoRepair's superior aggregate performance demonstrates that our documentation-enhanced approach is more broadly effective across the diverse software projects in SWE-bench Multimodal, where code-logic issues predominate over purely visual concerns.     

\begin{table}[tbp]
\caption{Repair Results (\%Resolved) Comparison Between RepoRepair, Agentless Lite and GUIRepair Across Repositories in SWE-bench Multimodal.}
\label{table3}
\centering
\begin{tabular}{c|ccc}
\toprule
Repository & Agentless Lite & GUIRepair & RepoRepair \\
\midrule
alibaba-fusion/next & 6/39 & 10/39 & 10/39 \\
bpmn-io/bpmn-js & 27/54 & 37/54 & 51/54 \\
carbon-design-system/carbon & 9/134 & 24/134 & 24/134 \\
eslint/eslint & 2/11 & 6/11 & 8/11 \\
GoogleChrome/lighthouse & 4/54 & 8/54 & 6/54 \\
grommet/grommet & 2/21 & 5/21 & 5/21 \\
highlightjs/highlight.js & 3/39 & 3/39 & 3/39 \\
openlayers/openlayers & 73/79 & 76/79 & 76/79 \\
prettier/prettier & 0/13 & 4/13 & 2/13 \\
PrismJS/prism & 5/38 & 8/38 & 4/38 \\
quarto-dev/quarto-cli & 0/24 & 5/24 & 3/24 \\
scratchfoundation/scratch-gui & 0/11 & 0/11 & 0/11 \\ \hline
\textbf{Total} & 131/517 (25.34\%) & 186/517 (35.98\%) & 192/517 (37.14\%)\\
\bottomrule
\end{tabular}
\end{table}


\textbf{Compared with baseline tools in Avg. \$Cost}. As shown in Table \ref{table1}, RepoRepair achieves a competitive repair rate of 45.67\% on SWE-bench Lite at an average cost of \$0.44 per fix, demonstrating significantly higher cost-efficiency than most agent-based approaches such as SWE-Agent (\$1.62 with Claude-3.5 Sonnet) and ExpeRepair (\$2.07 with Claude-3.5 Sonnet). On the more challenging SWE-bench Multimodal benchmark (Table \ref{table2}), although RepoRepair's cost is moderately higher than GUIRepair's most economical configuration (\$0.36 for 33.85\% with o4-mini), the \$0.20 additional cost is justified by the 3.28\% performance gain, offering better value for practical deployment where repair success is prioritized. Moreover, the cost efficiency of RepoRepair is further enhanced by its documentation reuse strategy. Specifically, on SWE-bench Lite, only \$0.12 is allocated to code documentation generation using DeepSeek-V3, while on SWE-bench Multimodal, this component increases to \$0.20 due to the larger scale and complexity of JavaScript/TypeScript repositories, with the remaining cost attributed to multimodal LLM inference using Claude-4 for localization and repair tasks.

\begin{table}[tbp]
\centering
\caption{File Localization Results (\%Correct Localization) Comparison Between RepoRepair and Agentless Lite Across Repositories in SWE-bench Lite. (AL = Agentless Lite; RR = RepoRepair)}
\label{table4}
\begin{tabular}{c|cc|cc}
\toprule
\multirow{2}{*}{Repository} & \multicolumn{2}{c}{File Retrieval} & \multicolumn{2}{c}{File Localization} \\ \cline{2-5} 
 & AL & RR & AL & RR \\
\midrule
astropy/astropy & 6/6 & 6/6 & 5/6 & 6/6 \\
django/django & 109/114 & 110/114 & 97/114 & 100/114 \\
matplotlib/matplotlib & 21/23 & 22/23 & 10/23 & 16/23 \\
mwaskom/seaborn & 4/4 & 4/4 & 2/4 & 3/4 \\
pallets/flask & 3/3 & 3/3 & 3/3 & 3/3 \\
psf/requests & 6/6 & 6/6 & 5/6 & 6/6 \\
pydata/xarray & 5/5 & 5/5 & 4/5 & 5/5 \\
pylint-dev/pylint & 6/6 & 4/6 & 6/6 & 3/6 \\
pytest-dev/pytest & 17/17 & 15/17 & 16/17 & 1/17 \\
scikit-learn/scikit-learn & 23/23 & 22/23 & 15/23 & 17/23 \\
sphinx-doc/sphinx & 16/16 & 16/16 & 12/16 & 12/16 \\
sympy/sympy & 73/77 & 69/77 & 49/77 & 65/77 \\ \hline
\textbf{Total} & 289/300 (96.33\%) & 282/300 (94\%) & 224/300 (74.67\%) & 237/300 (79\%) \\
\bottomrule
\end{tabular}
\end{table}

\begin{table}[tbp]
\centering
\caption{File Localization Results (\%Correct Localization) Comparison Between RepoRepair, Agentless Lite, and GUIRepair Across Repositories in SWE-bench Multimodal. (File R = File Retrieval; File L = File Localization; AL = Agentless Lite; GR = GUIRepair; RR = RepoRepair)}
\label{table5}
\begin{tabular}{c|cc|ccc}
\toprule
\multirow{2}{*}{Repository} & \multicolumn{2}{c}{File R} & \multicolumn{3}{c}{File L} \\ \cline{2-6} 
 & AL & RR & AL & GR & RR \\
\midrule
alibaba-fusion/next & 23/39 & 21/39 & 14/39 & 13/39 & 17/39 \\
bpmn-io/bpmn-js & 49/54 & 51/54 & 15/54 & 22/54 & 31/54 \\
carbon-design-system/carbon & 119/134 & 116/134 & 32/134 & 67/134 & 97/134 \\
eslint/eslint & 9/11 & 10/11 & 5/11 & 6/11 & 10/11 \\
GoogleChrome/lighthouse & 37/54 & 39/54 & 21/54 & 14/54 & 27/54 \\
grommet/grommet & 16/21 & 15/21 & 6/21 & 8/21 & 14/21 \\
highlightjs/highlight.js & 39/39 & 37/39 & 29/39 & 31/39 & 33/39 \\
openlayers/openlayers & 57/79 & 59/79 & 24/79 & 25/79 & 45/79 \\
prettier/prettier & 10/13 & 11/13 & 3/13 & 4/13 & 8/13 \\
PrismJS/prism & 37/38 & 24/38 & 8/38 & 24/38 & 16/38 \\
quarto-dev/quarto-cli & 11/24 & 12/24 & 0/24 & 15/24 & 9/24 \\
scratchfoundation/scratch-gui & 1/11 & 6/11 & 0/11 & 0/11 & 2/11 \\ \hline
\textbf{Total} & \begin{tabular}[c]{@{}c@{}}401/517 \\ (77.56\%)\end{tabular} & \begin{tabular}[c]{@{}c@{}}401/517 \\ (77.56\%)\end{tabular} & \begin{tabular}[c]{@{}c@{}}157/517 \\ (30.37\%)\end{tabular} & \begin{tabular}[c]{@{}c@{}}229/517 \\ (44.29\%)\end{tabular} & \begin{tabular}[c]{@{}c@{}}309/517 \\ (59.77\%)\end{tabular} \\
\bottomrule
\end{tabular}
\end{table}

\begin{table}[tbp]
\centering
\caption{Function-Level Localization Analysis of RepoRepair on two benchmarks.}
\label{table6}
\begin{tabular}{l|cc}
\toprule
Metric & SWE-bench Lite & SWE-bench Multimodal \\
\midrule
Successful file localizations & 237 & 309 \\
Successful function localizations & 205 (86.5\%) & 273 (88.35\%) \\
\vspace{0.1cm} \\
Function localization in 1 round & 222 (93.67\%) & 302 (97.73\%) \\
Function localization requiring >1 round & 15 (6.33\%) & 7 (2.27\%) \\
Average function localization rounds & 1.06 & 1.02 \\
\bottomrule
\end{tabular}
\end{table}

\textbf{Compared with baseline tools in \%Correct Localization}. In practical software development, accurately identifying modification locations provides significant debugging value beyond direct bug fixing. Note that while multiple valid modification locations may exist for a given issue (potentially differing from developer-chosen patches), comparison with developer patches remains a meaningful approximation metric for evaluating localization effectiveness. Although RepoRepair, Agentless Lite, and GUIRepair employ distinct localization workflows, all three approaches share the fundamental component of file localization (with both Agentless Lite and RepoRepair additionally incorporating similar file retrieval steps). This commonality enables a meaningful comparison of their file localization accuracy (\%Correct Localization), as quantitatively evaluated in Table \ref{table4} for SWE-bench Lite and Table \ref{table5} for SWE-bench Multimodal across different repositories.

On SWE-bench Lite (Table \ref{table4}), RepoRepair achieves comparable file retrieval accuracy (282/300, 94.0\%) to Agentless Lite (289/300, 96.3\%), while demonstrating superior file localization performance (237/300, 79.0\%) compared to Agentless Lite (224/300, 74.7\%). This performance gap in file retrieval may stem from our documentation-enhanced retrieval approach: RepoRepair directly matches issue descriptions with generated code documentation, while issue descriptions often contain code snippets and variable names that may not align perfectly with the natural language abstractions in code documentation, creating a semantic gap that affects retrieval accuracy. In contrast, Agentless Lite retrieves files based directly on code content, which may more accurately capture files relevant to the code-specific features mentioned in issue descriptions. Despite this minor disadvantage in the file retrieval stage, RepoRepair maintains superior performance in the subsequent file localization, indicating that RepoRepair provides stronger capability in correctly identifying suspicious files requiring modifications, thereby achieving overall improved localization performance.

On the more challenging SWE-bench Multimodal benchmark (Table \ref{table5}), RepoRepair demonstrates significant advantages in both file retrieval and localization. In file retrieval, RepoRepair achieves equal accuracy (401/517, 77.6\%) to Agentless Lite (401/517, 77.6\%), while substantially outperforming both baseline tools in file localization accuracy (309/517, 59.8\%) compared to Agentless Lite (157/517, 30.4\%) and GUIRepair (229/517, 44.3\%). For instance, in the 'carbon-design-system/carbon' repository, RepoRepair correctly localizes buggy files for 65 more issues than Agentless Lite and 30 more issues than GUIRepair. While this comparison might initially appear unfair due to differing methodologies (RepoRepair's Top-5 file localization vs. Agentless Lite's Top-1 approach), our evaluation framework actually favors Agentless Lite by employing a more lenient success criterion, requiring only that its single predicted file appears in the developer's patch set. In contrast, RepoRepair must correctly identify all files modified in the developer's patch to be counted as correct. This strict evaluation protocol confirms that the observed performance difference is both statistically significant and methodologically fair.

While Agentless Lite omits function-level localization, RepoRepair's hierarchical structure provides critical refinement by enabling LLMs to validate potentially relevant files further. Specifically, if no functions/classes are identified as relevant during this step, the file is excluded from repair, even if it was initially selected in file localization. This process essentially performs rigorous verification of the file-level results, preventing unnecessary repairs. Our evaluation applies strict criteria: any file where the function localization misses relevant functions/classes is considered a localization failure, regardless of subsequent repair outcomes. As shown in Table \ref{table6}, although function localization in RepoRepair excluded valid repair targets in 32 cases on SWE-bench Lite (from 237 to 205) and 36 cases on SWE-bench Multimodal (from 309 to 273), it achieved precision of 86.5\% (205/237) and 88.4\% (273/309) respectively, confirming its effectiveness as a precision improvement step in both benchmarks. Furthermore, function localization operates with high efficiency, completing 93.7\% of cases in a single round on SWE-bench Lite and 97.7\% on SWE-bench Multimodal, with average localization rounds of only 1.06 and 1.02, respectively.

\subsection{Ablation Study}

We conduct systematic ablation studies to evaluate the individual contributions of RepoRepair's key components and to validate its design rationale. (1) We assess the "file retrieval" component by disabling it during localization while retaining documentation-based file and function localization. (2) We evaluate the "code documentation" component by excluding it during localization (while keeping file retrieval), adopting Agentless's path-based approach instead. Notably, for SWE-bench Multimodal (JavaScript/TypeScript repositories with abundant anonymous functions), omitting documentation inherently excludes these critical elements; we skip function localization entirely and proceed directly to repair, whereas for SWE-bench Lite (Python repositories with minimal anonymous functions), we maintain the full hierarchical localization workflow. (3) We analyze the "code pruning" component by removing the context truncation step while preserving full code contexts. (4) To specifically address documentation quality concerns, we conduct a focused study on 'django/django' (114 bugs) from SWE-bench Lite, regenerating both file-level and function-level documentation using Qwen2.5-7B (a weaker model compared to DeepSeek-V3) and executing the complete localization-repair pipeline. Additionally, we perform human evaluation by randomly sampling 100 files with their corresponding code documentation to assess quality.

\begin{table}[tbp]
\centering
\caption{Computational Cost (Tokens) of RepoRepair Variants in Localization and Repair on two benchmarks. (File R = File Retrieval; File L = File Localization; Func L = Function Localization; CP = Code Pruning)}
\label{table7}
\begin{tabular}{c|cccc|cccc}
\toprule
\multicolumn{1}{l|}{} & \multicolumn{4}{c}{SWE-bench Lite} & \multicolumn{4}{c}{SWE-bench Multimodal} \\ \cline{2-9}
 & Full & w/o FR & w/o Doc & w/o CP & Full & w/o FR & w/o Doc & w/o CP \\  
\midrule
File R & 0 & - & 0 & 0 & 1.4M & - & 1.4M & 1.4M \\
File L & 2.9M & 44.6M & 0.6M & 2.9M & 23.5M & 132.6M & 2.9M & 23.5M \\
Func L & 6.3M & 13.5M & 0.3M & 6.3M & 6.2M & 16.5M & - & 6.2M \\
Repair & 2.2M & 5.4M & 1.9M & 3.8M & 8.1M & 21.6M & 9.0M & 12.5M \\ \hline
\textbf{Total} & 11.4M & 63.5M & 2.8M & 13.0M & 39.2M & 170.7M & 13.4M & 43.6M \\
\bottomrule
\end{tabular}
\end{table}

\begin{table}[tbp]
\caption{Performance of RepoRepair Variants in Localization (\%Correct Localization) and Repair (\%Resolved) on two benchmarks. (File R = File Retrieval; File L = File Localization; Func L = Function Localization; CP = Code Pruning)}
\label{table8}
\centering
\begin{tabular}{c|cccc|cccc}
\toprule
\multicolumn{1}{l|}{} & \multicolumn{4}{c|}{SWE-bench Lite} & \multicolumn{4}{c}{SWE-bench Multimodal} \\ \cline{2-9} 
 & Full & w/o FR & w/o Doc & w/o CP & Full & w/o FR & w/o Doc & w/o CP \\  
\midrule
File R & 282/300 & - & 282/300 & 282/300 & 401/517 & - & 401/517 & 401/517 \\
File L & 237/300 & 229/300 & 225/300 & 237/300 & 309/517 & 282/517 & 180/517 & 309/517 \\
Func L & 205/300 & 190/300 & 159/300 & 205/300 & 273/517 & 249/517 & - & 273/517 \\
Repair & 137/300 & 124/300 & 111/300 & 128/300 & 192/517 & 183/517 & 136/517 & 174/517\\
\bottomrule
\end{tabular}
\end{table}

\textbf{Effect of File Retrieval}. As shown in Table \ref{table7}, disabling file retrieval significantly increases computational costs across both benchmarks: on SWE-bench Lite, file localization token usage surges from 2.9M to 44.6M (a 15.4× increase), with total costs rising from 11.4M to 63.5M; on SWE-bench Multimodal, file localization increases from 23.5M to 132.6M, and total costs grow from 39.2M to 170.7M. This exponential cost escalation is expected, as repositories typically contain hundreds to thousands of files, and without preliminary filtering, the localization phase must process documentation for all files simultaneously. Table \ref{table8} further reveals that despite processing more information, the variant without file retrieval exhibits degraded performance: on SWE-bench Lite, file localization accuracy decreases from 237/300 to 229/300, and repair rate drops from 137/300 to 124/300; on SWE-bench Multimodal, file localization drops from 309/517 to 282/517, and repair decreases from 192/517 to 183/517. These results demonstrate that excessive context length degrades LLM reasoning capability, while file retrieval provides essential noise reduction, improving localization accuracy while reducing token costs by 82.1\% on Lite and 82.3\% on Multimodal—a crucial advantage for practical deployment.

\textbf{Effect of Code Documentation}. As shown in Table \ref{table7}, removing code documentation substantially reduces token usage during localization phases: on SWE-bench Lite, file localization costs decrease from 2.9M to 0.6M (a 79.3\% reduction), and total costs drop from 11.4M to 2.8M; on SWE-bench Multimodal, file localization decreases from 23.5M to 2.9M (an 87.7\% reduction), with total costs falling from 39.2M to 13.4M. However, Table \ref{table8} reveals a critical performance trade-off: without code documentation, localization accuracy suffers dramatically—on SWE-bench Lite, file localization drops from 237/300 to 225/300, and repair rate declines from 137/300 to 111/300; the impact is even more severe on SWE-bench Multimodal, where file localization plunges from 309/517 to 180/517, and repair falls from 192/517 to 136/517. This sharp decline indicates that while code documentation increases computational costs, it provides LLMs with essential semantic understanding that transcends surface-level code analysis. The documentation's structured representation of functionality, dependencies, and relationships enables more accurate identification of relevant code elements, which is particularly crucial for JavaScript/TypeScript repositories with complex anonymous functions and callbacks. The performance-cost analysis confirms that code documentation is indispensable for achieving practical repair effectiveness.

\textbf{Effect of Code Pruning}. Table \ref{table7} shows that disabling code pruning moderately increases computational costs, particularly in the repair phase: on SWE-bench Lite, repair token usage rises from 2.2M to 3.8M (a 72.7\% increase), and total costs increase from 11.4M to 13.0M; on SWE-bench Multimodal, repair costs grow from 8.1M to 12.5M (a 54.3\% increase), with total costs rising from 39.2M to 43.6M. This increase stems from providing LLMs with complete, unpruned code contexts during patch generation. Table \ref{table8} reveals an important performance pattern: on SWE-bench Lite, code pruning ablation maintains identical file and function localization accuracy (237/300 and 205/300 respectively) but reduces repair rate from 137/300 to 128/300; on SWE-bench Multimodal, it similarly preserves localization accuracy (309/517 and 273/517) but lowers repair rate from 192/517 to 174/517. These results indicate that while full code contexts provide comprehensive information, they also introduce substantial noise that can distract LLMs from focusing on the core bug-fixing logic. The code pruning component effectively filters irrelevant details, enabling more precise patch generation without sacrificing localization accuracy. This demonstrates that strategic context reduction—rather than maximal context provision—optimizes the repair process, particularly for complex bugs requiring focused attention on specific code segments.

\begin{table}[tbp]
\centering
\caption{Impact of Documentation Quality on RepoRepair's Performance (django/django). (File R = File Retrieval; File L = File Localization; Func L = Function Localization)}
\label{table9}
\small
\begin{tabular}{c|cccc}
\toprule
Doc Generator & File R & File L & Func L & Repair \\
\midrule
DeepSeek-V3 & 110/114 (96.5\%) & 100/114 (87.7\%) & 93/114 (81.6\%) & 67/114 (58.8\%) \\
Qwen2.5-7B & 98/114 (86.0\%) & 82/114 (71.9\%) & 68/114 (59.6\%) & 39/114 (34.2\%) \\
\bottomrule
\end{tabular}
\end{table}

\begin{table}[tbp]
\centering
\caption{Human Comparative Evaluation of Generated Documentation Quality (django/django).}
\label{table10}
\small
\begin{tabular}{c|ccc}
\toprule
Document Granularity & Sample Pairs & Prefer DeepSeek-V3 & Prefer Qwen2.5-7B \\
\midrule
File-level & 100 & 96 (96\%) & 4 (4\%) \\
Function-level & 627 & 615 (98.1\%) & 12 (1.9\%) \\
\bottomrule
\end{tabular}
\end{table}

\textbf{Effect of Documentation Quality}. To investigate how documentation generation capabilities affect RepoRepair's performance, we conduct a focused ablation study using two different LLMs for documentation generation: DeepSeek-V3 and Qwen2.5-7B. As shown in Table \ref{table9}, using Qwen2.5-7B for documentation generation results in noticeable performance degradation throughout the localization pipeline on django/django: file retrieval accuracy decreases from 96.5\% to 86.0\%, file localization drops from 87.7\% to 71.9\%, and function localization declines from 81.6\% to 59.6\%. This cascading degradation in localization performance directly leads to the substantial reduction in repair rate from 58.8\% to 34.2\%, demonstrating that documentation quality critically determines the success of the localization process, which in turn governs the upper bound of repair effectiveness.

To understand the underlying causes of this localization performance gap, we conduct a human comparative evaluation in which two authors with extensive Python development experience (both with over 4 years of research experience) assess documentation quality blindly. For each of 100 randomly sampled files and their corresponding 627 functions/classes, the evaluators independently compare anonymized documentation pairs from both models, selecting which version better describes the code's functionality based on three criteria: (1) accuracy of functional description, (2) completeness of key elements covered, and (3) clarity of expression. The evaluation followed a rigorous two-phase protocol. In the first phase, both evaluators independently assessed all samples. Their initial judgments agreed on 92\% of file-level pairs and 96.2\% of function-level pairs. For the remaining cases with disagreement, we implemented a resolution protocol: (a) first, the evaluators jointly reviewed the conflicting cases, discussing their rationale; (b) if consensus remained elusive, they consulted the source code more deeply; (c) as a final arbiter for 8 persistently disputed cases, they used GPT-4o to provide an independent analysis (while maintaining blindness to documentation origin), then reconciled their judgments based on this additional perspective. This multi-step resolution process ensured that all final judgments in Table \ref{table10} represent well-considered consensus.

As Table \ref{table10} shows, DeepSeek-V3-generated documentation is overwhelmingly preferred: 96\% at the file-level and 98.1\% at the function-level, compared to only 4\% and 1.9\% for Qwen2.5-7B, respectively. The near-universal preference for DeepSeek-V3 documentation indicates that the documentation generated by Qwen2.5-7B likely contains inaccuracies, omissions, or less precise semantic representations that directly impair the LLM's ability to accurately identify relevant code elements during localization. Since repair in RepoRepair operates strictly on the localized code segments, any deficiencies in documentation that hinder correct localization inevitably prevent successful repair. This finding underscores that high-quality documentation is essential for accurate localization—the foundational stage of our approach—and validates our design choice of leveraging advanced LLMs to generate precise semantic representations of repository code.

\subsection{Case Study}

\begin{figure*}[t!]
\centerline{\includegraphics[width=\textwidth]{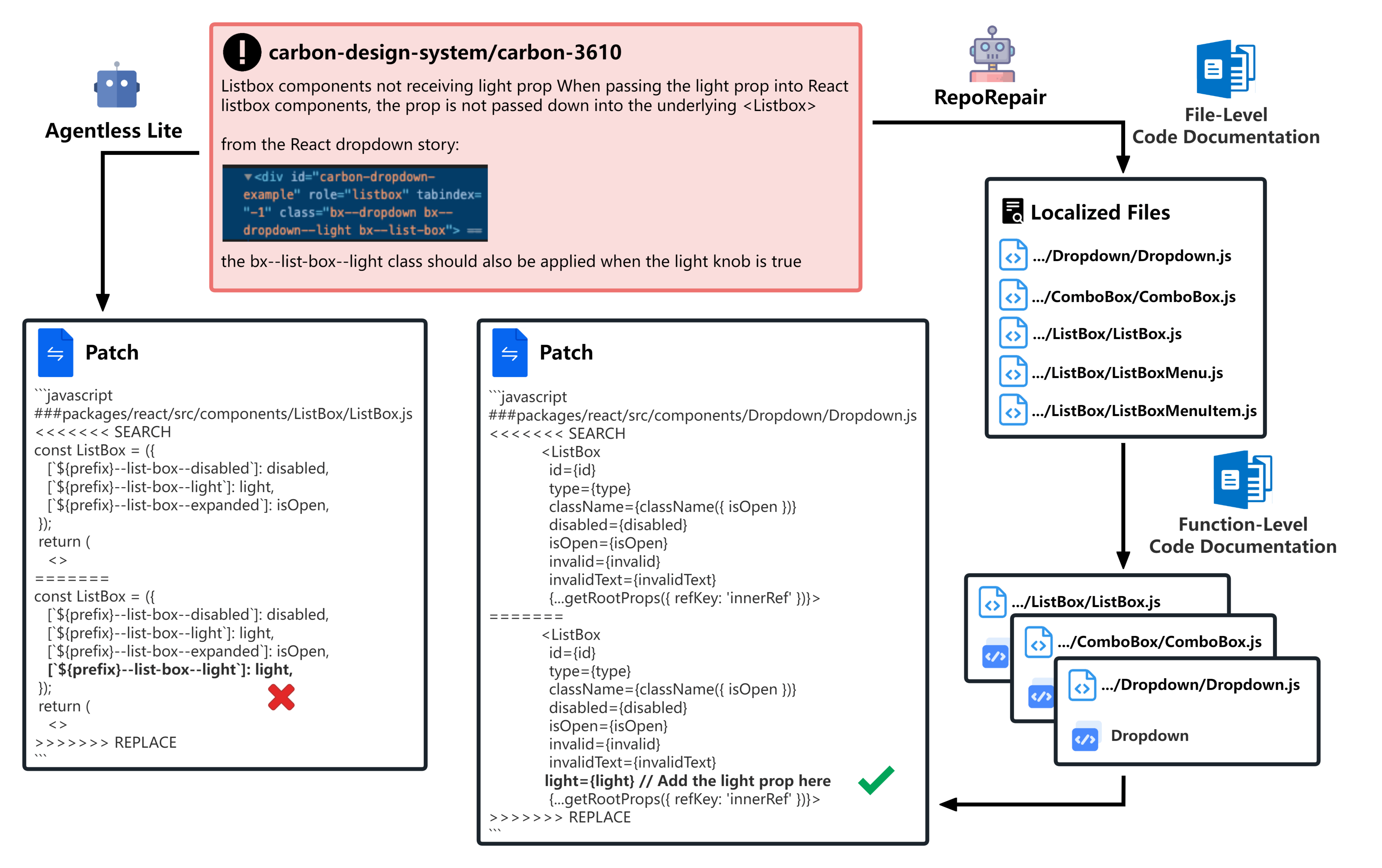}}
\caption{Case Comparison: Patches Generated by RepoRepair and Agentless Lite for Issue carbon-design-system/carbon-3610.}
\label{figure6}
\end{figure*}

To gain deeper insights into the effectiveness of RepoRepair's documentation-enhanced approach, we examine a representative case study from SWE-bench Multimodal. Figure \ref{figure6} presents the case 'carbon-design-system/carbon-3610', where the issue description reports a missing light prop propagation from React listbox components. Surface-level keyword matching (e.g., "listbox") would suggest modifying "ListBox.js"—as incorrectly done by baseline methods relying on code-based localization. In contrast, RepoRepair's hierarchical localization process, powered by code documentation, correctly identified the root cause in a different file: improper property passing in the higher-level component "Dropdown.js". The process involved: (1) initially selecting five suspicious files using file-level documentation, then (2) refining this to three files via function-level analysis, ultimately generating a correct patch that matches the developer's fix.

This case exemplifies RepoRepair's core strength: the ability to localize rarely mentioned but critical files by establishing semantic connections between issue descriptions and implementation code through comprehensive documentation. Without such structured documentation, LLMs struggle to move beyond surface keywords to understand the actual code relationships, leading to both localization and repair failures. The baseline's incorrect patch in "ListBox.js" (merely adding a CSS class) highlights the limitations of keyword-driven approaches and underscores the value of RepoRepair's documentation-aware reasoning for complex, repository-wide issues.

\section{Threats to Validity}
\label{sec:threats}

\textbf{Internal Validity}. A primary internal validity threat concerns potential data leakage. Although Claude-4 was released after the construction of SWE-bench Lite, it demonstrates robust generalization on the more recent SWE-bench Multimodal benchmark, mitigating temporal data contamination concerns. To rigorously examine leakage risks, we perform the following checks: (1) removing all ground-truth contextual information from model inputs, and (2) verifying that the model never outputs correct files, functions, or code snippets without appropriate retrieved context. Our analysis reveals minor leakage in SWE-bench Lite, limited to 4 instances in file localization where the model identified correct file paths despite retrieval failures—possibly due to exposure to similar public repositories during pre-training. No such leakage was detected in SWE-bench Multimodal. Overall, we conclude that benchmark contamination is well-controlled and does not substantially affect the validity of our results.

Another threat stems from model selection. We use Claude-4 consistently across comparisons, as it is widely adopted by competitive baselines (e.g., SWE-Agent, ExpeRepair) on these benchmarks. Our ablation studies confirm that RepoRepair's performance gains stem primarily from its documentation mechanism rather than the underlying model choice.

\textbf{External Validity}. The main external threats relate to language coverage and repository scale. Our evaluation spans two major language ecosystems (Python, JavaScript/TypeScript) across SWE-bench Lite and SWE-bench Multimodal, yet performance in other popular languages (e.g., Java, C\#) remains unvalidated. Additionally, while the benchmarks represent real-world projects, they may not capture the scale or unique patterns of some industrial codebases where computational limits could emerge. Nonetheless, our multi-benchmark results across distinct language domains demonstrate promising generalizability for RepoRepair.

\section{Limitations}
\label{sec:limitations}

RepoRepair’s design exhibits several limitations that constrain its broader applicability in repository-level program repair:

A fundamental limitation stems from its code documentation generation mechanism: since it exclusively processes functions/classes in code files, files containing no functional components (e.g., configuration files, static asset bundles) cannot generate documentation. When such files are critical for issue resolution, RepoRepair is likely to fail. Our analysis identifies 26 such cases (4.2\%) in SWE-bench Multimodal, primarily involving configuration files (e.g., prism-sql.js), static style files (e.g., style.css), and entry files (e.g., index.js). Consequently, RepoRepair exhibits significantly reduced effectiveness for non-functional changes (e.g., parameter tuning, style changes) compared to functional bug fixes. In practical deployment, this limitation can be addressed through complementary techniques. For instance, configuration files could be processed using specialized parsers to extract structured information, and maintaining a repository-wide function call graph would enable more targeted documentation updates when related files change, avoiding complete repository re-documentation.

Furthermore, the workflow implemented by RepoRepair is predicated on the assumption that defects are confined to modifications within pre-existing files. As a result, the tool is unable to accommodate feature requests that necessitate the creation of entirely new files—a limitation that impacts 13 cases (2.1\%) in SWE-bench Multimodal. This constraint is not unique to RepoRepair but reflects a fundamental challenge shared across agent-free APR tools: the inability to distinguish between issue types (modification vs. creation) and adapt localization and repair workflows accordingly. Future extensions could incorporate a file creation phase when no existing files are deemed relevant, prompting the LLM to generate appropriate new file structures based on repository patterns.

Additionally, our incremental documentation generation strategy reuses documentation for unchanged files between commits to improve efficiency. However, this approach may fail to capture how an unchanged file's role or dependencies evolve as surrounding code changes—a subtle contextual shift that could affect localization accuracy. Maintaining lightweight dependency tracking would enable more precise documentation updates, regenerating documentation not only for changed files but also for files whose architectural relationships have evolved.

\section{Conclusion and Future Works}
\label{sec:conclusion}

We propose RepoRepair, a novel approach for automatically repairing repository-level issues by leveraging code documentation. Following the agent-free methodology, RepoRepair performs repairs through three key stages: hierarchical documentation generation, multimodal fault localization, and context-aware patch generation. Comprehensive evaluation on both SWE-bench Lite and the latest SWE-bench Multimodal benchmark demonstrates that our approach achieves state-of-the-art performance, with repair rates of 45.7\% and 37.1\% respectively, while maintaining high cost-efficiency at \$0.44 and \$0.56 per fix. Ablation studies and detailed analysis confirm that code documentation significantly enhances both localization accuracy and repair effectiveness, particularly for complex, cross-file issues requiring deep repository understanding.

As to future works, we plan to extend RepoRepair along two directions to mitigate existing limitations: (1) enhancing documentation coverage to diverse file types beyond functions/classes, including configuration files and static assets, and (2) supporting additive changes through extended workflows or controlled agent-based methods for handling feature requests requiring file creation. These improvements aim to broaden RepoRepair's applicability and robustness across diverse real-world development scenarios while maintaining its core advantages in comprehension and cost-effectiveness.



\begin{acks}
This research is supported by the National Natural Science Foundation of China (62572235), CCF-Huawei Populus Grove Fund, and Cooperation Fund of Huawei-NJU Creative Laboratory for Novel Software Technology.
\end{acks}

\bibliographystyle{ACM-Reference-Format}
\bibliography{mybib}

@String{Computer = "{IEEE} Computer" }

@article{gpt4,
  author       = {OpenAI},
  title        = {{GPT-4} Technical Report},
  journal      = {CoRR},
  volume       = {abs/2303.08774},
  year         = {2023},
  url          = {https://doi.org/10.48550/arXiv.2303.08774},
  doi          = {10.48550/ARXIV.2303.08774},
  eprinttype    = {arXiv},
  eprint       = {2303.08774},
  bibsource    = {dblp computer science bibliography, https://dblp.org}
}

@article{deepseek,
  author       = {DeepSeek{-}AI and
                  Aixin Liu and
                  Bei Feng and
                  Bing Xue and
                  Bingxuan Wang and
                  Bochao Wu and
                  Chengda Lu and
                  Chenggang Zhao and
                  Chengqi Deng and
                  Chenyu Zhang and
                  Chong Ruan and
                  Damai Dai and
                  Daya Guo and
                  Dejian Yang and
                  Deli Chen and
                  Dongjie Ji and
                  Erhang Li and
                  Fangyun Lin and
                  Fucong Dai and
                  Fuli Luo and
                  Guangbo Hao and
                  Guanting Chen and
                  Guowei Li and
                  H. Zhang and
                  Han Bao and
                  Hanwei Xu and
                  Haocheng Wang and
                  Haowei Zhang and
                  Honghui Ding and
                  Huajian Xin and
                  Huazuo Gao and
                  Hui Li and
                  Hui Qu and
                  J. L. Cai and
                  Jian Liang and
                  Jianzhong Guo and
                  Jiaqi Ni and
                  Jiashi Li and
                  Jiawei Wang and
                  Jin Chen and
                  Jingchang Chen and
                  Jingyang Yuan and
                  Junjie Qiu and
                  Junlong Li and
                  Junxiao Song and
                  Kai Dong and
                  Kai Hu and
                  Kaige Gao and
                  Kang Guan and
                  Kexin Huang and
                  Kuai Yu and
                  Lean Wang and
                  Lecong Zhang and
                  Lei Xu and
                  Leyi Xia and
                  Liang Zhao and
                  Litong Wang and
                  Liyue Zhang and
                  Meng Li and
                  Miaojun Wang and
                  Mingchuan Zhang and
                  Minghua Zhang and
                  Minghui Tang and
                  Mingming Li and
                  Ning Tian and
                  Panpan Huang and
                  Peiyi Wang and
                  Peng Zhang and
                  Qiancheng Wang and
                  Qihao Zhu and
                  Qinyu Chen and
                  Qiushi Du and
                  R. J. Chen and
                  R. L. Jin and
                  Ruiqi Ge and
                  Ruisong Zhang and
                  Ruizhe Pan and
                  Runji Wang and
                  Runxin Xu and
                  Ruoyu Zhang and
                  Ruyi Chen and
                  S. S. Li and
                  Shanghao Lu and
                  Shangyan Zhou and
                  Shanhuang Chen and
                  Shaoqing Wu and
                  Shengfeng Ye and
                  Shengfeng Ye and
                  Shirong Ma and
                  Shiyu Wang and
                  Shuang Zhou and
                  Shuiping Yu and
                  Shunfeng Zhou and
                  Shuting Pan and
                  T. Wang and
                  Tao Yun and
                  Tian Pei and
                  Tianyu Sun and
                  W. L. Xiao and
                  Wangding Zeng},
  title        = {DeepSeek-V3 Technical Report},
  journal      = {CoRR},
  volume       = {abs/2412.19437},
  year         = {2024},
  url          = {https://doi.org/10.48550/arXiv.2412.19437},
  doi          = {10.48550/ARXIV.2412.19437},
  eprinttype    = {arXiv},
  eprint       = {2412.19437},
  bibsource    = {dblp computer science bibliography, https://dblp.org}
}

@inproceedings{swe-bench,
  author       = {Carlos E. Jimenez and
                  John Yang and
                  Alexander Wettig and
                  Shunyu Yao and
                  Kexin Pei and
                  Ofir Press and
                  Karthik R. Narasimhan},
  title        = {SWE-bench: Can Language Models Resolve Real-world Github Issues?},
  booktitle    = {The Twelfth International Conference on Learning Representations,
                  {ICLR} 2024, Vienna, Austria, May 7-11, 2024},
  publisher    = {OpenReview.net},
  year         = {2024},
  url          = {https://openreview.net/forum?id=VTF8yNQM66},
  bibsource    = {dblp computer science bibliography, https://dblp.org}
}

@inproceedings{swe-bench-m,
  author       = {John Yang and
                  Carlos E. Jimenez and
                  Alex L. Zhang and
                  Kilian Lieret and
                  Joyce Yang and
                  Xindi Wu and
                  Ori Press and
                  Niklas Muennighoff and
                  Gabriel Synnaeve and
                  Karthik R. Narasimhan and
                  Diyi Yang and
                  Sida Wang and
                  Ofir Press},
  title        = {SWE-bench Multimodal: Do {AI} Systems Generalize to Visual Software
                  Domains?},
  booktitle    = {The Thirteenth International Conference on Learning Representations,
                  {ICLR} 2025, Singapore, April 24-28, 2025},
  publisher    = {OpenReview.net},
  year         = {2025},
  url          = {https://openreview.net/forum?id=riTiq3i21b},
  bibsource    = {dblp computer science bibliography, https://dblp.org}
}

@inproceedings{autocoderover,
  author       = {Yuntong Zhang and
                  Haifeng Ruan and
                  Zhiyu Fan and
                  Abhik Roychoudhury},
  editor       = {Maria Christakis and
                  Michael Pradel},
  title        = {AutoCodeRover: Autonomous Program Improvement},
  booktitle    = {Proceedings of the 33rd {ACM} {SIGSOFT} International Symposium on
                  Software Testing and Analysis, {ISSTA} 2024, Vienna, Austria, September
                  16-20, 2024},
  pages        = {1592--1604},
  publisher    = {{ACM}},
  year         = {2024},
  url          = {https://doi.org/10.1145/3650212.3680384},
  doi          = {10.1145/3650212.3680384},
  bibsource    = {dblp computer science bibliography, https://dblp.org}
}

@article{agentless,
  author       = {Chunqiu Steven Xia and
                  Yinlin Deng and
                  Soren Dunn and
                  Lingming Zhang},
  title        = {Agentless: Demystifying LLM-based Software Engineering Agents},
  journal      = {CoRR},
  volume       = {abs/2407.01489},
  year         = {2024},
  url          = {https://doi.org/10.48550/arXiv.2407.01489},
  doi          = {10.48550/ARXIV.2407.01489},
  eprinttype    = {arXiv},
  eprint       = {2407.01489},
  bibsource    = {dblp computer science bibliography, https://dblp.org}
}

@inproceedings{swe-agent,
  author       = {John Yang and
                  Carlos E. Jimenez and
                  Alexander Wettig and
                  Kilian Lieret and
                  Shunyu Yao and
                  Karthik Narasimhan and
                  Ofir Press},
  editor       = {Amir Globersons and
                  Lester Mackey and
                  Danielle Belgrave and
                  Angela Fan and
                  Ulrich Paquet and
                  Jakub M. Tomczak and
                  Cheng Zhang},
  title        = {SWE-agent: Agent-Computer Interfaces Enable Automated Software Engineering},
  booktitle    = {Advances in Neural Information Processing Systems 38: Annual Conference
                  on Neural Information Processing Systems 2024, NeurIPS 2024, Vancouver,
                  BC, Canada, December 10 - 15, 2024},
  year         = {2024},
  url          = {http://papers.nips.cc/paper\_files/paper/2024/hash/5a7c947568c1b1328ccc5230172e1e7c-Abstract-Conference.html},
  bibsource    = {dblp computer science bibliography, https://dblp.org}
}

@article{repoagent,
  author       = {Qinyu Luo and
                  Yining Ye and
                  Shihao Liang and
                  Zhong Zhang and
                  Yujia Qin and
                  Yaxi Lu and
                  Yesai Wu and
                  Xin Cong and
                  Yankai Lin and
                  Yingli Zhang and
                  Xiaoyin Che and
                  Zhiyuan Liu and
                  Maosong Sun},
  title        = {RepoAgent: An LLM-Powered Open-Source Framework for Repository-level
                  Code Documentation Generation},
  journal      = {CoRR},
  volume       = {abs/2402.16667},
  year         = {2024},
  url          = {https://doi.org/10.48550/arXiv.2402.16667},
  doi          = {10.48550/ARXIV.2402.16667},
  eprinttype    = {arXiv},
  eprint       = {2402.16667},
  bibsource    = {dblp computer science bibliography, https://dblp.org}
}

@article{code_generation_1,
  author       = {Yujia Li and
                  David H. Choi and
                  Junyoung Chung and
                  Nate Kushman and
                  Julian Schrittwieser and
                  R{\'{e}}mi Leblond and
                  Tom Eccles and
                  James Keeling and
                  Felix Gimeno and
                  Agustin Dal Lago and
                  Thomas Hubert and
                  Peter Choy and
                  Cyprien de Masson d'Autume and
                  Igor Babuschkin and
                  Xinyun Chen and
                  Po{-}Sen Huang and
                  Johannes Welbl and
                  Sven Gowal and
                  Alexey Cherepanov and
                  James Molloy and
                  Daniel J. Mankowitz and
                  Esme Sutherland Robson and
                  Pushmeet Kohli and
                  Nando de Freitas and
                  Koray Kavukcuoglu and
                  Oriol Vinyals},
  title        = {Competition-Level Code Generation with AlphaCode},
  journal      = {CoRR},
  volume       = {abs/2203.07814},
  year         = {2022},
  url          = {https://doi.org/10.48550/arXiv.2203.07814},
  doi          = {10.48550/ARXIV.2203.07814},
  eprinttype    = {arXiv},
  eprint       = {2203.07814},
  bibsource    = {dblp computer science bibliography, https://dblp.org}
}

@article{code_generation_2,
  author       = {Noah Patton and
                  Kia Rahmani and
                  Meghana Missula and
                  Joydeep Biswas and
                  Isil Dillig},
  title        = {Programming-by-Demonstration for Long-Horizon Robot Tasks},
  journal      = {Proc. {ACM} Program. Lang.},
  volume       = {8},
  number       = {{POPL}},
  pages        = {512--545},
  year         = {2024},
  url          = {https://doi.org/10.1145/3632860},
  doi          = {10.1145/3632860},
  bibsource    = {dblp computer science bibliography, https://dblp.org}
}

@article{code_generation_3,
  author       = {Mark Chen and
                  Jerry Tworek and
                  Heewoo Jun and
                  Qiming Yuan and
                  Henrique Pond{\'{e}} de Oliveira Pinto and
                  Jared Kaplan and
                  Harri Edwards and
                  Yuri Burda and
                  Nicholas Joseph and
                  Greg Brockman and
                  Alex Ray and
                  Raul Puri and
                  Gretchen Krueger and
                  Michael Petrov and
                  Heidy Khlaaf and
                  Girish Sastry and
                  Pamela Mishkin and
                  Brooke Chan and
                  Scott Gray and
                  Nick Ryder and
                  Mikhail Pavlov and
                  Alethea Power and
                  Lukasz Kaiser and
                  Mohammad Bavarian and
                  Clemens Winter and
                  Philippe Tillet and
                  Felipe Petroski Such and
                  Dave Cummings and
                  Matthias Plappert and
                  Fotios Chantzis and
                  Elizabeth Barnes and
                  Ariel Herbert{-}Voss and
                  William Hebgen Guss and
                  Alex Nichol and
                  Alex Paino and
                  Nikolas Tezak and
                  Jie Tang and
                  Igor Babuschkin and
                  Suchir Balaji and
                  Shantanu Jain and
                  William Saunders and
                  Christopher Hesse and
                  Andrew N. Carr and
                  Jan Leike and
                  Joshua Achiam and
                  Vedant Misra and
                  Evan Morikawa and
                  Alec Radford and
                  Matthew Knight and
                  Miles Brundage and
                  Mira Murati and
                  Katie Mayer and
                  Peter Welinder and
                  Bob McGrew and
                  Dario Amodei and
                  Sam McCandlish and
                  Ilya Sutskever and
                  Wojciech Zaremba},
  title        = {Evaluating Large Language Models Trained on Code},
  journal      = {CoRR},
  volume       = {abs/2107.03374},
  year         = {2021},
  url          = {https://arxiv.org/abs/2107.03374},
  eprinttype    = {arXiv},
  eprint       = {2107.03374},
  bibsource    = {dblp computer science bibliography, https://dblp.org}
}

@inproceedings{code_translation_1,
  author       = {Baptiste Rozi{\`{e}}re and
                  Jie Zhang and
                  Fran{\c{c}}ois Charton and
                  Mark Harman and
                  Gabriel Synnaeve and
                  Guillaume Lample},
  title        = {Leveraging Automated Unit Tests for Unsupervised Code Translation},
  booktitle    = {The Tenth International Conference on Learning Representations, {ICLR}
                  2022, Virtual Event, April 25-29, 2022},
  publisher    = {OpenReview.net},
  year         = {2022},
  url          = {https://openreview.net/forum?id=cmt-6KtR4c4},
  bibsource    = {dblp computer science bibliography, https://dblp.org}
}

@inproceedings{code_translation_2,
  author       = {Rangeet Pan and
                  Ali Reza Ibrahimzada and
                  Rahul Krishna and
                  Divya Sankar and
                  Lambert Pouguem Wassi and
                  Michele Merler and
                  Boris Sobolev and
                  Raju Pavuluri and
                  Saurabh Sinha and
                  Reyhaneh Jabbarvand},
  title        = {Lost in Translation: {A} Study of Bugs Introduced by Large Language
                  Models while Translating Code},
  booktitle    = {Proceedings of the 46th {IEEE/ACM} International Conference on Software
                  Engineering, {ICSE} 2024, Lisbon, Portugal, April 14-20, 2024},
  pages        = {82:1--82:13},
  publisher    = {{ACM}},
  year         = {2024},
  url          = {https://doi.org/10.1145/3597503.3639226},
  doi          = {10.1145/3597503.3639226},
  bibsource    = {dblp computer science bibliography, https://dblp.org}
}

@inproceedings{code_translation_3,
  author       = {Baptiste Rozi{\`{e}}re and
                  Marie{-}Anne Lachaux and
                  Lowik Chanussot and
                  Guillaume Lample},
  editor       = {Hugo Larochelle and
                  Marc'Aurelio Ranzato and
                  Raia Hadsell and
                  Maria{-}Florina Balcan and
                  Hsuan{-}Tien Lin},
  title        = {Unsupervised Translation of Programming Languages},
  booktitle    = {Advances in Neural Information Processing Systems 33: Annual Conference
                  on Neural Information Processing Systems 2020, NeurIPS 2020, December
                  6-12, 2020, virtual},
  year         = {2020},
  url          = {https://proceedings.neurips.cc/paper/2020/hash/ed23fbf18c2cd35f8c7f8de44f85c08d-Abstract.html},
  bibsource    = {dblp computer science bibliography, https://dblp.org}
}

@inproceedings{code_doc_1,
  author       = {Sergio Cozzetti B. de Souza and
                  Nicolas Anquetil and
                  K{\'{a}}thia Mar{\c{c}}al de Oliveira},
  editor       = {Scott R. Tilley and
                  Robert M. Newman},
  title        = {A study of the documentation essential to software maintenance},
  booktitle    = {Proceedings of the 23rd Annual International Conference on Design
                  of Communication: documenting {\&} Designing for Pervasive Information,
                  {SIGDOC} 2005, Coventry, UK, September 21-23, 2005},
  pages        = {68--75},
  publisher    = {{ACM}},
  year         = {2005},
  url          = {https://doi.org/10.1145/1085313.1085331},
  doi          = {10.1145/1085313.1085331},
  bibsource    = {dblp computer science bibliography, https://dblp.org}
}

@inproceedings{code_doc_2,
  author       = {Xin Xia and
                  Lingfeng Bao and
                  David Lo and
                  Zhenchang Xing and
                  Ahmed E. Hassan and
                  Shanping Li},
  editor       = {Michel Chaudron and
                  Ivica Crnkovic and
                  Marsha Chechik and
                  Mark Harman},
  title        = {Measuring program comprehension: a large-scale field study with professionals},
  booktitle    = {Proceedings of the 40th International Conference on Software Engineering,
                  {ICSE} 2018, Gothenburg, Sweden, May 27 - June 03, 2018},
  pages        = {584},
  publisher    = {{ACM}},
  year         = {2018},
  url          = {https://doi.org/10.1145/3180155.3182538},
  doi          = {10.1145/3180155.3182538},
  bibsource    = {dblp computer science bibliography, https://dblp.org}
}

@article{code_doc_3,
  author       = {Junji Zhi and
                  Vahid Garousi{-}Yusifoglu and
                  Bo Sun and
                  Golara Garousi and
                  S. M. Shahnewaz and
                  G{\"{u}}nther Ruhe},
  title        = {Cost, benefits and quality of software development documentation:
                  {A} systematic mapping},
  journal      = {J. Syst. Softw.},
  volume       = {99},
  pages        = {175--198},
  year         = {2015},
  url          = {https://doi.org/10.1016/j.jss.2014.09.042},
  doi          = {10.1016/J.JSS.2014.09.042},
  bibsource    = {dblp computer science bibliography, https://dblp.org}
}

@inproceedings{defects4j,
  author       = {Ren{\'{e}} Just and
                  Darioush Jalali and
                  Michael D. Ernst},
  editor       = {Corina S. Pasareanu and
                  Darko Marinov},
  title        = {Defects4J: a database of existing faults to enable controlled testing
                  studies for Java programs},
  booktitle    = {International Symposium on Software Testing and Analysis, {ISSTA}
                  '14, San Jose, CA, {USA} - July 21 - 26, 2014},
  pages        = {437--440},
  publisher    = {{ACM}},
  year         = {2014},
  url          = {https://doi.org/10.1145/2610384.2628055},
  doi          = {10.1145/2610384.2628055},
  bibsource    = {dblp computer science bibliography, https://dblp.org}
}

@inproceedings{alpharepair,
  author       = {Chunqiu Steven Xia and
                  Lingming Zhang},
  editor       = {Abhik Roychoudhury and
                  Cristian Cadar and
                  Miryung Kim},
  title        = {Less training, more repairing please: revisiting automated program
                  repair via zero-shot learning},
  booktitle    = {Proceedings of the 30th {ACM} Joint European Software Engineering
                  Conference and Symposium on the Foundations of Software Engineering,
                  {ESEC/FSE} 2022, Singapore, Singapore, November 14-18, 2022},
  pages        = {959--971},
  publisher    = {{ACM}},
  year         = {2022},
  url          = {https://doi.org/10.1145/3540250.3549101},
  doi          = {10.1145/3540250.3549101},
  bibsource    = {dblp computer science bibliography, https://dblp.org}
}

@article{repairllama,
  author       = {Andr{\'{e}} Silva and
                  Sen Fang and
                  Martin Monperrus},
  title        = {RepairLLaMA: Efficient Representations and Fine-Tuned Adapters for
                  Program Repair},
  journal      = {CoRR},
  volume       = {abs/2312.15698},
  year         = {2023},
  url          = {https://doi.org/10.48550/arXiv.2312.15698},
  doi          = {10.48550/ARXIV.2312.15698},
  eprinttype    = {arXiv},
  eprint       = {2312.15698},
  bibsource    = {dblp computer science bibliography, https://dblp.org}
}

@article{coder,
  author       = {Dong Chen and
                  Shaoxin Lin and
                  Muhan Zeng and
                  Daoguang Zan and
                  Jian{-}Gang Wang and
                  Anton Cheshkov and
                  Jun Sun and
                  Hao Yu and
                  Guoliang Dong and
                  Artem Aliev and
                  Jie Wang and
                  Xiao Cheng and
                  Guangtai Liang and
                  Yuchi Ma and
                  Pan Bian and
                  Tao Xie and
                  Qianxiang Wang},
  title        = {CodeR: Issue Resolving with Multi-Agent and Task Graphs},
  journal      = {CoRR},
  volume       = {abs/2406.01304},
  year         = {2024},
  url          = {https://doi.org/10.48550/arXiv.2406.01304},
  doi          = {10.48550/ARXIV.2406.01304},
  eprinttype    = {arXiv},
  eprint       = {2406.01304},
  bibsource    = {dblp computer science bibliography, https://dblp.org}
}

@inproceedings{swe-search,
  author       = {Antonis Antoniades and
                  Albert {\"{O}}rwall and
                  Kexun Zhang and
                  Yuxi Xie and
                  Anirudh Goyal and
                  William Yang Wang},
  title        = {SWE-Search: Enhancing Software Agents with Monte Carlo Tree Search
                  and Iterative Refinement},
  booktitle    = {The Thirteenth International Conference on Learning Representations,
                  {ICLR} 2025, Singapore, April 24-28, 2025},
  publisher    = {OpenReview.net},
  year         = {2025},
  url          = {https://openreview.net/forum?id=G7sIFXugTX},
  bibsource    = {dblp computer science bibliography, https://dblp.org}
}

@article{mcts,
  author       = {David Silver and
                  Aja Huang and
                  Chris J. Maddison and
                  Arthur Guez and
                  Laurent Sifre and
                  George van den Driessche and
                  Julian Schrittwieser and
                  Ioannis Antonoglou and
                  Vedavyas Panneershelvam and
                  Marc Lanctot and
                  Sander Dieleman and
                  Dominik Grewe and
                  John Nham and
                  Nal Kalchbrenner and
                  Ilya Sutskever and
                  Timothy P. Lillicrap and
                  Madeleine Leach and
                  Koray Kavukcuoglu and
                  Thore Graepel and
                  Demis Hassabis},
  title        = {Mastering the game of Go with deep neural networks and tree search},
  journal      = {Nat.},
  volume       = {529},
  number       = {7587},
  pages        = {484--489},
  year         = {2016},
  url          = {https://doi.org/10.1038/nature16961},
  doi          = {10.1038/NATURE16961},
  bibsource    = {dblp computer science bibliography, https://dblp.org}
}

@inproceedings{doc_generation_1,
  author       = {Srinivasan Iyer and
                  Ioannis Konstas and
                  Alvin Cheung and
                  Luke Zettlemoyer},
  title        = {Summarizing Source Code using a Neural Attention Model},
  booktitle    = {Proceedings of the 54th Annual Meeting of the Association for Computational
                  Linguistics, {ACL} 2016, August 7-12, 2016, Berlin, Germany, Volume
                  1: Long Papers},
  publisher    = {The Association for Computer Linguistics},
  year         = {2016},
  url          = {https://doi.org/10.18653/v1/p16-1195},
  doi          = {10.18653/V1/P16-1195},
  bibsource    = {dblp computer science bibliography, https://dblp.org}
}

@inproceedings{doc_generation_2,
  author       = {Miltiadis Allamanis and
                  Hao Peng and
                  Charles Sutton},
  editor       = {Maria{-}Florina Balcan and
                  Kilian Q. Weinberger},
  title        = {A Convolutional Attention Network for Extreme Summarization of Source
                  Code},
  booktitle    = {Proceedings of the 33nd International Conference on Machine Learning,
                  {ICML} 2016, New York City, NY, USA, June 19-24, 2016},
  series       = {{JMLR} Workshop and Conference Proceedings},
  volume       = {48},
  pages        = {2091--2100},
  publisher    = {JMLR.org},
  year         = {2016},
  url          = {http://proceedings.mlr.press/v48/allamanis16.html},
  bibsource    = {dblp computer science bibliography, https://dblp.org}
}

@inproceedings{doc_generation_3,
  author       = {Ashish Vaswani and
                  Noam Shazeer and
                  Niki Parmar and
                  Jakob Uszkoreit and
                  Llion Jones and
                  Aidan N. Gomez and
                  Lukasz Kaiser and
                  Illia Polosukhin},
  editor       = {Isabelle Guyon and
                  Ulrike von Luxburg and
                  Samy Bengio and
                  Hanna M. Wallach and
                  Rob Fergus and
                  S. V. N. Vishwanathan and
                  Roman Garnett},
  title        = {Attention is All you Need},
  booktitle    = {Advances in Neural Information Processing Systems 30: Annual Conference
                  on Neural Information Processing Systems 2017, December 4-9, 2017,
                  Long Beach, CA, {USA}},
  pages        = {5998--6008},
  year         = {2017},
  url          = {https://proceedings.neurips.cc/paper/2017/hash/3f5ee243547dee91fbd053c1c4a845aa-Abstract.html},
  bibsource    = {dblp computer science bibliography, https://dblp.org}
}

@inproceedings{doc_generation_4,
  author       = {Zhangyin Feng and
                  Daya Guo and
                  Duyu Tang and
                  Nan Duan and
                  Xiaocheng Feng and
                  Ming Gong and
                  Linjun Shou and
                  Bing Qin and
                  Ting Liu and
                  Daxin Jiang and
                  Ming Zhou},
  editor       = {Trevor Cohn and
                  Yulan He and
                  Yang Liu},
  title        = {CodeBERT: {A} Pre-Trained Model for Programming and Natural Languages},
  booktitle    = {Findings of the Association for Computational Linguistics: {EMNLP}
                  2020, Online Event, 16-20 November 2020},
  series       = {Findings of {ACL}},
  volume       = {{EMNLP} 2020},
  pages        = {1536--1547},
  publisher    = {Association for Computational Linguistics},
  year         = {2020},
  url          = {https://doi.org/10.18653/v1/2020.findings-emnlp.139},
  doi          = {10.18653/V1/2020.FINDINGS-EMNLP.139},
  bibsource    = {dblp computer science bibliography, https://dblp.org}
}

@inproceedings{doc_generation_5,
  author       = {Yue Wang and
                  Weishi Wang and
                  Shafiq R. Joty and
                  Steven C. H. Hoi},
  editor       = {Marie{-}Francine Moens and
                  Xuanjing Huang and
                  Lucia Specia and
                  Scott Wen{-}tau Yih},
  title        = {CodeT5: Identifier-aware Unified Pre-trained Encoder-Decoder Models
                  for Code Understanding and Generation},
  booktitle    = {Proceedings of the 2021 Conference on Empirical Methods in Natural
                  Language Processing, {EMNLP} 2021, Virtual Event / Punta Cana, Dominican
                  Republic, 7-11 November, 2021},
  pages        = {8696--8708},
  publisher    = {Association for Computational Linguistics},
  year         = {2021},
  url          = {https://doi.org/10.18653/v1/2021.emnlp-main.685},
  doi          = {10.18653/V1/2021.EMNLP-MAIN.685},
  bibsource    = {dblp computer science bibliography, https://dblp.org}
}

@inproceedings{doc_generation_6,
  author       = {Junaed Younus Khan and
                  Gias Uddin},
  title        = {Automatic Code Documentation Generation Using {GPT-3}},
  booktitle    = {37th {IEEE/ACM} International Conference on Automated Software Engineering,
                  {ASE} 2022, Rochester, MI, USA, October 10-14, 2022},
  pages        = {174:1--174:6},
  publisher    = {{ACM}},
  year         = {2022},
  url          = {https://doi.org/10.1145/3551349.3559548},
  doi          = {10.1145/3551349.3559548},
  bibsource    = {dblp computer science bibliography, https://dblp.org}
}

@article{docagent,
  author       = {Dayu Yang and
                  Antoine Simoulin and
                  Xin Qian and
                  Xiaoyi Liu and
                  Yuwei Cao and
                  Zhaopu Teng and
                  Grey Yang},
  title        = {DocAgent: {A} Multi-Agent System for Automated Code Documentation
                  Generation},
  journal      = {CoRR},
  volume       = {abs/2504.08725},
  year         = {2025},
  url          = {https://doi.org/10.48550/arXiv.2504.08725},
  doi          = {10.48550/ARXIV.2504.08725},
  eprinttype    = {arXiv},
  eprint       = {2504.08725},
  bibsource    = {dblp computer science bibliography, https://dblp.org}
}

@article{faiss,
  author       = {Jeff Johnson and
                  Matthijs Douze and
                  Herv{\'{e}} J{\'{e}}gou},
  title        = {Billion-Scale Similarity Search with GPUs},
  journal      = {{IEEE} Trans. Big Data},
  volume       = {7},
  number       = {3},
  pages        = {535--547},
  year         = {2021},
  url          = {https://doi.org/10.1109/TBDATA.2019.2921572},
  doi          = {10.1109/TBDATA.2019.2921572},
  bibsource    = {dblp computer science bibliography, https://dblp.org}
}

@inproceedings{cot,
  author       = {Jason Wei and
                  Xuezhi Wang and
                  Dale Schuurmans and
                  Maarten Bosma and
                  Brian Ichter and
                  Fei Xia and
                  Ed H. Chi and
                  Quoc V. Le and
                  Denny Zhou},
  editor       = {Sanmi Koyejo and
                  S. Mohamed and
                  A. Agarwal and
                  Danielle Belgrave and
                  K. Cho and
                  A. Oh},
  title        = {Chain-of-Thought Prompting Elicits Reasoning in Large Language Models},
  booktitle    = {Advances in Neural Information Processing Systems 35: Annual Conference
                  on Neural Information Processing Systems 2022, NeurIPS 2022, New Orleans,
                  LA, USA, November 28 - December 9, 2022},
  year         = {2022},
  url          = {http://papers.nips.cc/paper\_files/paper/2022/hash/9d5609613524ecf4f15af0f7b31abca4-Abstract-Conference.html},
  bibsource    = {dblp computer science bibliography, https://dblp.org}
}

@inproceedings{test,
  author       = {W. Eric Wong and
                  Joseph R. Horgan and
                  Saul London and
                  Hiralal Agrawal},
  title        = {A study of effective regression testing in practice},
  booktitle    = {Eighth International Symposium on Software Reliability Engineering,
                  {ISSRE} 1997, Albuquerque, NM, USA, November 2-5, 1997},
  pages        = {264--274},
  publisher    = {{IEEE} Computer Society},
  year         = {1997},
  url          = {https://doi.org/10.1109/ISSRE.1997.630875},
  doi          = {10.1109/ISSRE.1997.630875},
  bibsource    = {dblp computer science bibliography, https://dblp.org}
}

@inproceedings{repair-1,
  author       = {Chunqiu Steven Xia and
                  Yuxiang Wei and
                  Lingming Zhang},
  title        = {Automated Program Repair in the Era of Large Pre-trained Language
                  Models},
  booktitle    = {45th {IEEE/ACM} International Conference on Software Engineering,
                  {ICSE} 2023, Melbourne, Australia, May 14-20, 2023},
  pages        = {1482--1494},
  publisher    = {{IEEE}},
  year         = {2023},
  url          = {https://doi.org/10.1109/ICSE48619.2023.00129},
  doi          = {10.1109/ICSE48619.2023.00129},
  bibsource    = {dblp computer science bibliography, https://dblp.org}
}

@article{repair-2,
  author       = {Chunqiu Steven Xia and
                  Lingming Zhang},
  title        = {Keep the Conversation Going: Fixing 162 out of 337 bugs for {\textdollar}0.42
                  each using ChatGPT},
  journal      = {CoRR},
  volume       = {abs/2304.00385},
  year         = {2023},
  url          = {https://doi.org/10.48550/arXiv.2304.00385},
  doi          = {10.48550/ARXIV.2304.00385},
  eprinttype    = {arXiv},
  eprint       = {2304.00385},
  bibsource    = {dblp computer science bibliography, https://dblp.org}
}

@article{repair-3,
  author       = {Islem Bouzenia and
                  Premkumar T. Devanbu and
                  Michael Pradel},
  title        = {RepairAgent: An Autonomous, LLM-Based Agent for Program Repair},
  journal      = {CoRR},
  volume       = {abs/2403.17134},
  year         = {2024},
  url          = {https://doi.org/10.48550/arXiv.2403.17134},
  doi          = {10.48550/ARXIV.2403.17134},
  eprinttype    = {arXiv},
  eprint       = {2403.17134},
  bibsource    = {dblp computer science bibliography, https://dblp.org}
}

@inproceedings{openhands,
  author       = {Xingyao Wang and
                  Boxuan Li and
                  Yufan Song and
                  Frank F. Xu and
                  Xiangru Tang and
                  Mingchen Zhuge and
                  Jiayi Pan and
                  Yueqi Song and
                  Bowen Li and
                  Jaskirat Singh and
                  Hoang H. Tran and
                  Fuqiang Li and
                  Ren Ma and
                  Mingzhang Zheng and
                  Bill Qian and
                  Yanjun Shao and
                  Niklas Muennighoff and
                  Yizhe Zhang and
                  Binyuan Hui and
                  Junyang Lin and
                  et al.},
  title        = {OpenHands: An Open Platform for {AI} Software Developers as Generalist
                  Agents},
  booktitle    = {The Thirteenth International Conference on Learning Representations,
                  {ICLR} 2025, Singapore, April 24-28, 2025},
  publisher    = {OpenReview.net},
  year         = {2025},
  url          = {https://openreview.net/forum?id=OJd3ayDDoF},
  bibsource    = {dblp computer science bibliography, https://dblp.org}
}

@article{dars,
  author       = {Vaibhav Aggarwal and
                  Ojasv Kamal and
                  Abhinav Japesh and
                  Zhijing Jin and
                  Bernhard Sch{\"{o}}lkopf},
  title        = {{DARS:} Dynamic Action Re-Sampling to Enhance Coding Agent Performance
                  by Adaptive Tree Traversal},
  journal      = {CoRR},
  volume       = {abs/2503.14269},
  year         = {2025},
  url          = {https://doi.org/10.48550/arXiv.2503.14269},
  doi          = {10.48550/ARXIV.2503.14269},
  eprinttype    = {arXiv},
  eprint       = {2503.14269},
  bibsource    = {dblp computer science bibliography, https://dblp.org}
}

@article{ssim,
  author       = {Zhou Wang and
                  Alan C. Bovik and
                  Hamid R. Sheikh and
                  Eero P. Simoncelli},
  title        = {Image quality assessment: from error visibility to structural similarity},
  journal      = {{IEEE} Trans. Image Process.},
  volume       = {13},
  number       = {4},
  pages        = {600--612},
  year         = {2004},
  url          = {https://doi.org/10.1109/TIP.2003.819861},
  doi          = {10.1109/TIP.2003.819861},
  bibsource    = {dblp computer science bibliography, https://dblp.org}
}

@inproceedings{retrival-2,
  author       = {Haotian Liu and
                  Chunyuan Li and
                  Qingyang Wu and
                  Yong Jae Lee},
  editor       = {Alice Oh and
                  Tristan Naumann and
                  Amir Globerson and
                  Kate Saenko and
                  Moritz Hardt and
                  Sergey Levine},
  title        = {Visual Instruction Tuning},
  booktitle    = {Advances in Neural Information Processing Systems 36: Annual Conference
                  on Neural Information Processing Systems 2023, NeurIPS 2023, New Orleans,
                  LA, USA, December 10 - 16, 2023},
  year         = {2023},
  url          = {http://papers.nips.cc/paper\_files/paper/2023/hash/6dcf277ea32ce3288914faf369fe6de0-Abstract-Conference.html},
  bibsource    = {dblp computer science bibliography, https://dblp.org}
}

@inproceedings{clip,
  author       = {Alec Radford and
                  Jong Wook Kim and
                  Chris Hallacy and
                  Aditya Ramesh and
                  Gabriel Goh and
                  Sandhini Agarwal and
                  Girish Sastry and
                  Amanda Askell and
                  Pamela Mishkin and
                  Jack Clark and
                  Gretchen Krueger and
                  Ilya Sutskever},
  editor       = {Marina Meila and
                  Tong Zhang},
  title        = {Learning Transferable Visual Models From Natural Language Supervision},
  booktitle    = {Proceedings of the 38th International Conference on Machine Learning,
                  {ICML} 2021, 18-24 July 2021, Virtual Event},
  series       = {Proceedings of Machine Learning Research},
  volume       = {139},
  pages        = {8748--8763},
  publisher    = {{PMLR}},
  year         = {2021},
  url          = {http://proceedings.mlr.press/v139/radford21a.html},
  bibsource    = {dblp computer science bibliography, https://dblp.org}
}

@article{openhands-2,
  author       = {Aditya Bharat Soni and
                  Boxuan Li and
                  Xingyao Wang and
                  Valerie Chen and
                  Graham Neubig},
  title        = {Coding Agents with Multimodal Browsing are Generalist Problem Solvers},
  journal      = {CoRR},
  volume       = {abs/2506.03011},
  year         = {2025},
  url          = {https://doi.org/10.48550/arXiv.2506.03011},
  doi          = {10.48550/ARXIV.2506.03011},
  eprinttype    = {arXiv},
  eprint       = {2506.03011},
  bibsource    = {dblp computer science bibliography, https://dblp.org}
}

@article{experepair,
  author       = {Fangwen Mu and
                  Junjie Wang and
                  Lin Shi and
                  Song Wang and
                  Shoubin Li and
                  Qing Wang},
  title        = {{EXPEREPAIR:} Dual-Memory Enhanced LLM-based Repository-Level Program
                  Repair},
  journal      = {CoRR},
  volume       = {abs/2506.10484},
  year         = {2025},
  url          = {https://doi.org/10.48550/arXiv.2506.10484},
  doi          = {10.48550/ARXIV.2506.10484},
  eprinttype    = {arXiv},
  eprint       = {2506.10484},
  bibsource    = {dblp computer science bibliography, https://dblp.org}
}

@article{guirepair,
  author       = {Kai Huang and
                  Jian Zhang and
                  Xiaofei Xie and
                  Chunyang Chen},
  title        = {Seeing is Fixing: Cross-Modal Reasoning with Multimodal LLMs for Visual
                  Software Issue Fixing},
  journal      = {CoRR},
  volume       = {abs/2506.16136},
  year         = {2025},
  url          = {https://doi.org/10.48550/arXiv.2506.16136},
  doi          = {10.48550/ARXIV.2506.16136},
  eprinttype    = {arXiv},
  eprint       = {2506.16136},
  bibsource    = {dblp computer science bibliography, https://dblp.org}
}

@article{patchpilot,
  author       = {Hongwei Li and
                  Yuheng Tang and
                  Shiqi Wang and
                  Wenbo Guo},
  title        = {PatchPilot: {A} Stable and Cost-Efficient Agentic Patching Framework},
  journal      = {CoRR},
  volume       = {abs/2502.02747},
  year         = {2025},
  url          = {https://doi.org/10.48550/arXiv.2502.02747},
  doi          = {10.48550/ARXIV.2502.02747},
  eprinttype    = {arXiv},
  eprint       = {2502.02747},
  bibsource    = {dblp computer science bibliography, https://dblp.org}
}

@inproceedings{specrover,
  author       = {Haifeng Ruan and
                  Yuntong Zhang and
                  Abhik Roychoudhury},
  title        = {SpecRover: Code Intent Extraction via LLMs},
  booktitle    = {47th {IEEE/ACM} International Conference on Software Engineering,
                  {ICSE} 2025, Ottawa, ON, Canada, April 26 - May 6, 2025},
  pages        = {963--974},
  publisher    = {{IEEE}},
  year         = {2025},
  url          = {https://doi.org/10.1109/ICSE55347.2025.00080},
  doi          = {10.1109/ICSE55347.2025.00080},
  bibsource    = {dblp computer science bibliography, https://dblp.org}
}

@inproceedings{llmao,
  author       = {Aidan Z. H. Yang and
                  Claire {Le Goues} and
                  Ruben Martins and
                  Vincent J. Hellendoorn},
  title        = {Large Language Models for Test-Free Fault Localization},
  booktitle    = {Proceedings of the 46th {IEEE/ACM} International Conference on Software
                  Engineering, {ICSE} 2024, Lisbon, Portugal, April 14-20, 2024},
  pages        = {17:1--17:12},
  publisher    = {{ACM}},
  year         = {2024},
  url          = {https://doi.org/10.1145/3597503.3623342},
  doi          = {10.1145/3597503.3623342},
  bibsource    = {dblp computer science bibliography, https://dblp.org}
}

@inproceedings{fusefl,
  author       = {Ratnadira Widyasari and
                  Jia Wei Ang and
                  Truong Giang Nguyen and
                  Neil Sharma and
                  David Lo},
  title        = {Demystifying Faulty Code: Step-by-Step Reasoning for Explainable Fault
                  Localization},
  booktitle    = {{IEEE} International Conference on Software Analysis, Evolution and
                  Reengineering, {SANER} 2024, Rovaniemi, Finland, March 12-15, 2024},
  pages        = {568--579},
  publisher    = {{IEEE}},
  year         = {2024},
  url          = {https://doi.org/10.1109/SANER60148.2024.00064},
  doi          = {10.1109/SANER60148.2024.00064},
  bibsource    = {dblp computer science bibliography, https://dblp.org}
}

@inproceedings{rlce,
  author       = {Yuxiao Chen and
                  Jingzheng Wu and
                  Xiang Ling and
                  Changjiang Li and
                  Zhiqing Rui and
                  Tianyue Luo and
                  Yanjun Wu},
  title        = {When Large Language Models Confront Repository-Level Automatic Program
                  Repair: How Well They Done?},
  booktitle    = {Proceedings of the 2024 {IEEE/ACM} 46th International Conference on
                  Software Engineering: Companion Proceedings, {ICSE} Companion 2024,
                  Lisbon, Portugal, April 14-20, 2024},
  pages        = {459--471},
  publisher    = {{ACM}},
  year         = {2024},
  url          = {https://doi.org/10.1145/3639478.3647633},
  doi          = {10.1145/3639478.3647633},
  bibsource    = {dblp computer science bibliography, https://dblp.org}
}

\end{document}